\documentclass[12pt]{article} 
\usepackage{amsfonts} 					
\usepackage{amssymb} 					
\usepackage{amsmath} 					
\usepackage{geometry} 					
\usepackage{fancyhdr} 					
\usepackage{graphicx} 					
\usepackage{epsf} 						
\usepackage{epsfig} 					
\usepackage[bottom]{footmisc}			
\usepackage{natbib} 					
\usepackage{setspace}					
\usepackage{rotating}					
\usepackage{threeparttable} 	
\usepackage{booktabs}					
\usepackage{caption}					
\usepackage{array} 						
\usepackage{float} 						
\usepackage{forest}						
\usepackage{tikz} 						
\usepackage{pgfkeys}
\usepackage{comment}					
\usepackage{xcolor}						
\usepackage{multirow}					
\usepackage{mathtools}					
\usepackage{pdflscape} 					
\usepackage{subcaption}					

\usepackage[colorlinks, urlcolor=blue, linkcolor=blue,citecolor=blue,backref=page]{hyperref}

\usepackage{float}
\usepackage{amsmath,amsfonts,esint}
\usepackage{amsmath,amstext,rotating}
\usepackage{amsfonts,amssymb,graphics,xspace,endnotes}
\usepackage{lineno}
\usepackage{amsthm}
\usepackage{physics}
\usepackage{graphicx}
\usepackage{verbatim}
\usepackage{natbib}
\usepackage{comment}
\usepackage{algorithm}
\usepackage{algorithmic}
\usepackage[utf8]{inputenc} 
\usepackage[T1]{fontenc} 
\usepackage{pgfplots}
\pgfplotsset{compat=1.18}
\usepackage{booktabs,threeparttable,float}

\theoremstyle{plain}

\numberwithin{equation}{section}

\theoremstyle{remark}

\usetikzlibrary{patterns}

\usepackage{natbib} 
\setcitestyle{sort&compress}

\AtBeginDocument{}

\hypersetup{
    colorlinks,
    linkcolor={blue!75!black},
    urlcolor={blue!75!black},
    citecolor={blue!75!black},
}

\title{Understanding Classical Decomposability of Inequality Measures: A Graphical Analysis}

\author{
Tatiana Komarova\thanks{Corresponding author. Faculty of Economics, University of Cambridge.  tk670@cam.ac.uk.}} 

\date{April 16, 2026 \\ \vspace{.5cm}}

\begin{document}
\maketitle

\spacing{1}

\begin{abstract} 

This paper develops a geometric diagnostic framework for classical
inequality decomposability.  Representing the simplest nontrivial setting of three-person income
distributions as points on the two-dimensional income-share simplex,
we translate population-share-weighted and income-share-weighted
decomposability into concrete geometric restrictions on within- and
between-group residuals, making it possible to localise and
characterise violations across measures.  Applied to the Mean Log
Deviation, the Gini coefficient, the coefficient of variation, and
the Theil index, the analysis shows that decomposability is not a
binary property as measures fail in qualitatively distinct ways, and
the between-group residual is consistently the primary locus of
failure.  Negative between-group residuals render the
decomposition uninterpretable and arise for the coefficient of
variation and the Theil index under population-share weighting, and
for the Mean Log Deviation under income-share weighting.  Stylised
numerical examples quantify the resulting misinterpretation scenarios for
applied researchers.
\vspace{0.05in}

\begin{description}
\item[Keywords:] Inequality indices, decomposability, simplex geometry

\end{description}
\end{abstract}

\spacing{1.19}

\section{Introduction} 

Empirical studies of income inequality routinely decompose total
inequality into within-group and between-group components.  This
framing is standard in both academic and policy work, including
global assessments of inequality dynamics where within- and
between-country components are used to gauge the relative importance
of subgroup differences \citep{WorldInequalityReport2022}.  The
between-group component is conventionally interpreted as measuring
inequality attributable to differences in subgroup means, while the
within-group component captures residual inequality that would remain
if those mean differences were eliminated.

A substantial theoretical literature has examined when such
decompositions are valid.  Under standard axioms, only a small class
of measures admits exact additive decompositions: population-share
weighting uniquely characterises the Mean Log Deviation, while
income-share weighting characterises the Theil index
\citep{Bourguignon1979,Shorrocks1980}.  Widely used measures such as
the Gini coefficient do not generally satisfy these properties,
implying that their within- and between-group components need not
admit a structural interpretation.  The applied literature has
responded in part by adopting alternative decomposition methods that
guarantee additivity for arbitrary measures, most prominently through
Shapley-value-based attribution rules \citep{Shorrocks2013,
CowellFiorio2011}.  These methods are increasingly common in
empirical work, though the resulting components correspond to average
marginal effects rather than structural within- and between-group
inequality. For textbook treatments of inequality measurement and decomposition, see \citet{Cowell1995} and \citet{Jenkins1991}.

Despite this literature, the structure of decomposability failures remains only partially characterized. Existing contributions typically establish whether a given measure admits an exact decomposition, often through axiomatic or functional restrictions, but provide limited guidance on where along the income distribution violations arise or how large they are in economically relevant regions. This paper develops tools to characterize the location and magnitude of decomposability failures and to quantify the risks faced by practitioners who apply classical decompositions.

The central analytical objects of this paper are two composite residuals. Given a partition of the population into subgroups, any inequality measure $I$ can be decomposed in two complementary ways. First,  one can subtract from total inequality the between-group component implied by classical decomposability, leaving a within-group residual $g_1=I-B$. Second,  one can subtract the within-group component, leaving a between-group residual $g_2=I-W$. Under exact decomposability, $g_1$ equals the weighted sum of within-group inequalities $W$ and $g_2$ equals the between-group term $B$, so both have clean structural interpretations. When decomposability fails, studying $g_1$ and $g_2$ simultaneously is more informative than studying the single composite residual $R = I -W - B$ that has been the focus of the prior literature. As the paper shows, these two residuals do not fail symmetrically.  
This asymmetry is invisible when attention is restricted to the single residual $R$.

This paper proposes a geometric diagnostic framework for diagnosing classical decomposability properties and their violations on the basis of two residuals.  It is the paper's main contribution and it is developed for a population of $n=3$ individuals as this is the smallest population for
which within- and between-group variation is simultaneously
non-trivial.\footnote{For smaller populations every inequality measure
satisfies both decomposability notions trivially (Section \ref{sec:setting}).}  For
$n=3$, scale invariance allows income distributions to be represented
as points on the two-dimensional simplex of income shares.  Each
decomposability requirement translates into testable geometric
restrictions on the within- and between-group residuals, which are an
invariance property, a Schur-convexity property,
and sign and identification conditions.  The
simplex representation converts abstract axiomatic conditions into
visual, verifiable geometric restrictions.

Three main findings emerge. First, decomposability is not a binary property. The four measures examined depart from classical additivity in qualitatively distinct ways, localised in distinct regions of the simplex. Second, the between-group residual $g_2$ is consistently the primary locus of failure. Across all measures and weighting schemes examined, the within-group residual $g_1$ preserves natural Schur-convexity and non-negativity properties, which is not the case for the between-group residual $g_2$. Third, $g_2$ can turn negative for certain income distributions. A negative $g_2$ is a strictly stronger pathology than a large or poorly-signed composite residual $R=I-W-B$ as it means the sign of the entire between-group attribution is inverted, so the decomposition actively attributes negative inequality to cross-group mean differences. These failures  arise for the coefficient of variation and the Theil index under population-share weighting, and for the Mean Log Deviation under income-share weighting. The failures happen at empirically plausible distributions, as Section \ref{sec:misinterpretation} demonstrates with stylised numerical examples.

A further contribution is that the geometric framework extends
beyond the two classical weighting schemes.  Section \ref{sec:extensions}
shows how the simplex diagnostics apply to any decomposability notion
of the standard additive form, and illustrates this using the recent
axiomatic decomposition of the Gini coefficient by
\cite{HeikkuriSchief2026}, whose within-group component corresponds
to a geometrically distinct invariance principle.  The simplex
framework thereby serves as a common diagnostic language for
comparing classical and non-classical decompositions.

The paper builds on and extends several strands of the prior literature. The difficulties of decomposing rank-based measures were identified early. \citet{Pyatt1976} showed that the Gini coefficient admits a three-component decomposition into within-group, between-group, and an overlap residual whenever subgroup income ranges are not perfectly separated, and \citet{MookherjeeShorrocks1982} characterised that residual as a quantity that is structurally necessary to maintain the decomposition identity but resists economic interpretation. \citet{LambertAronson1993} subsequently gave the residual a geometric reading as a sub-area of the Lorenz diagram, and \citet{LambertDecoster2005} showed that it reflects the Gini’s sensitivity to income overlap rather than a defect of any particular decomposition formula. The present paper’s contribution relative to this strand is to shift attention from the single residual $R = I - W - B$ to the two composite residuals $g_1$ and $g_2$ studied separately, to embed the analysis in the simplex geometry of income-share distributions, and to apply the resulting framework uniformly across four widely used measures and both classical weighting schemes.

In the applied literature this paper  relates to e.g. \citet{ElbersLanjouw2008} that documents that
between-group estimates can diverge in sign as well as magnitude from
the true between-group inequality when group distributions
overlap. \citet{MorduchSicular2002} show that alternative
decomposition methods applied to the same data can yield qualitatively
different pictures of within- versus between-group inequality.  Both
of these applied findings are consistent with the between-group residuals properties identified analytically in this paper. 

The contribution of this paper thus lies not in establishing that many inequality 
measures fail classical decomposability (this is known)  but in providing a unified
geometric language that makes the {structure} of each failure
transparent. The graphical approach of the paper can be seen as complementing recent extensions like path-independent decompositions (\cite{FOSTER2000}) and inequality analyses in \cite{HeikkuriSchief2026}.

\section{Setting and decomposability}
\label{sec:setting}

Consider inequality measures that are symmetric, differentiable, homogeneous of degree zero,
strictly Schur--convex, and normalized to zero under equality.

Following \cite{Bourguignon1979}, denote $I^q(x_1, x_2,\ldots ,x_q)$ as the inequality measure with such properties for a population of $q$ individuals with incomes $x_1, x_2,\ldots ,x_q$.  

The two classical decomposability properties this paper  focuses on are \textit{population-share-weighted} and the \textit{income-share-weighted} decomposabilities. The first one   is defined as 
\begin{equation}
\label{eq:bourguignon11}
\tag{$DCP$}
I^n(y_1,...,y_n)
=
\sum_{i=1}^{m} \frac{n_i}{n}\,
I^{n_i}\!\left(y_{i1}, y_{i2}, \ldots, y_{i n_i}\right)
+
I^n\!\left(\underbrace{\bar y_1, \ldots, \bar y_1}_{n_1 \text{ times}}, \underbrace{\bar y_2, \ldots,  \bar y_2}_{n_2 \text{ times}}, \ldots, \underbrace{\bar y_m, \ldots, \bar y_m}_{n_m \text{ times}}\right),
\end{equation} 
and the second one is defined as 
\begin{equation}
\label{eq:bourguignon12}
\tag{$DCI$}
I^n(y_1,...,y_n)
=
\sum_{i=1}^{m} \frac{Y_i}{Y}\,
I^{n_i}\!\left(y_{i1}, y_{i2}, \ldots, y_{i n_i}\right)
+
I^n\!\left(\underbrace{\bar y_1, \ldots, \bar y_1}_{n_1 \text{ times}}, \underbrace{\bar y_2, \ldots,  \bar y_2}_{n_2 \text{ times}}, \ldots, \underbrace{\bar y_m, \ldots, \bar y_m}_{n_m \text{ times}}\right),
\end{equation} 
where $n $ is the total number of individuals, $m$  is the number of subgroups, $n_i$  is the size of $i$'s subgroup, $\sum_{i=1}^{m} n_i = n$,
 $y_{ij}$  is the income of individual  $j$  in subgroup  $i$,
 $\bar y_i $ is the mean income of subgroup  $i$, $Y_i = \sum_{j=1}^{n_i} y_{ij}$  is the total income of subgroup $i$,  and $Y$ is the total income of the whole population: $Y=\sum_{i=1}^{m} Y_i$.

The paper focuses on  a simple setting of the population of size 3. $n=3$ is the canonical non-trivial setting for decomposability as for $n \leq 2$, every inequality measure satisfies both \eqref{eq:bourguignon11} and \eqref{eq:bourguignon12} trivially, since within-group inequality vanishes in any singleton subgroup and the between-group term accounts for all measured inequality: for any inequality measure, 
$$I^{2}(y_{1}, y_{2})=\underbrace{\frac{1}{2} I^{1}(y_{1})+\frac{1}{2} I^{1}(y_{2})}_{=0 \; \text{ within}} + \underbrace{I^{2}(y_{1}, y_{2})}_{\text{ between}}$$
for population-share-weighted decomposability, and
$$I^{2}(y_{1}, y_{2})=\underbrace{\frac{y_1}{y_1+y_2} I^{1}(y_{1})+\frac{y_2}{y_1+y_2} I^{1}( y_{2})}_{=0 \; \text{ within}} + \underbrace{I^{2}(y_{1}, y_{2})}_{\text{ between}}$$ 
for income-share-weighted decomposability. 
The three-person population is therefore the smallest for which within- and between-group variation are simultaneously non-trivial, making it the natural focus for a geometric analysis.

Let a population of three individuals have incomes $(y_1,y_2,y_3)$ such that $y_1+y_2+y_3 > 0$, and define income shares
\[
z_i = \frac{y_i}{y_1+y_2+y_3}, \quad i=1,2,3.
\]
The vector $z=(z_1,z_2,z_3)$ lies on the simplex
\[
\Delta_2 = \{ z \in \mathbb{R}^3_+ : z_1+z_2+z_3=1 \}.
\]

The only nontrivial population partition in this setting consists of a subgroup of two individuals
and a subgroup of one individual.

\textit{Population-share-weighted decomposability} requires
\begin{equation}
I^3(z_1,z_2,z_3)
=
\frac{2}{3} I^2(z_1,z_2) + \frac{1}{3}\underbrace{I^1(z_3)}_{=0}
+
I^3\!\left(\frac{z_1+z_2}{2},\frac{z_1+z_2}{2}, z_3\right), 
\label{eq:DCP}
\end{equation}
so to analyze this property define the \textit{within-group residual} $g_1$ and the \textit{between-group residual} $g_2$, respectively, as 
\begin{align*}
g_1(z)
&=
I^3(z_1,z_2,z_3)
-
I^3\!\left(\frac{z_1+z_2}{2},\frac{z_1+z_2}{2}, z_3\right),\\
g_2(z)
&=
I^3(z_1,z_2,z_3)
-
\frac{2}{3} I^2(z_1,z_2).
\end{align*}

\textit{Income-share-weighted decomposability} requires 
\begin{equation}
 I^3(z_1,z_2,z_3)=(z_1+z_2) I^2(z_1,z_2) + z_3\underbrace{I^1(z_3)}_{=0} + I^3(\frac{z_1+z_2}{2}, \frac{z_1+z_2}{2}, z_3), 
\label{eq:DCI}
\end{equation}
so to analyze this property define the \textit{within-group residual} $g_1^{\mathcal{I}}$ and the \textit{between-group residual} $g_2^{\mathcal{I}}$, respectively, as 
\begin{align*} g_1^{\mathcal{I}}(z) & := \frac{I^3(z) -I^3(\frac{z_1+z_2}{2}, \frac{z_1+z_2}{2}, z_3)}{z_1+z_2}, \quad z_1+z_2>0\\
g_2^{\mathcal{I}}(z) & := I^3(z) -(z_1+z_2) I^2(z_1,z_2) .
\end{align*} 

Under exact relevant decomposability, $g_1$ ($g_1^{\mathcal{I}}$) captures within-group inequality
and $g_2$ ($g_2^{\mathcal{I}}$) captures between-group inequality. 

One might worry that the simplex analysis relies on a special low-dimensional case of the three–person population. As all inequality measures considered here are symmetric and homogeneous of degree zero, higher-dimensional behavior must reduce locally to configurations analogous to those represented on the two-dimensional simplex. In this sense, the three–person simplex captures the essential geometry of decomposability. Since larger-population decompositions contain many embedded comparisons of the same basic $2+1$ type, violations observed in the three-person case should be understood as evidence of a structural incompatibility between the index and the decomposition rule, rather than as a peculiarity of low dimension. 

Take any $n$ and consider the share distribution across this population to be $$ (z_1,\, z_2,\, \underbrace{\frac{z_3}{n-2},\ldots,\frac{z_3}{n-2}}_{n-2}),$$
where $z_3=1-z_1-z_2$. Consider  the population as partitioned into the subgroup $\{z_1,z_2\}$ and the subgroup of $n-2$ members each holding 
equal income share $z_3/(n-2)$, so that the total income share of the second subgroup 
is $z_3$ throughout. The failure of population-share-weighted decomposability means that 
\begin{equation*}
    I^n\!\left(z_1,z_2,\frac{z_3}{n-2},\ldots,\frac{z_3}{n-2}\right) 
    \neq \frac{2}{n}I^2(z_1,z_2) + 
    I^n\!\left(\frac{z_1+z_2}{2},\frac{z_1+z_2}{2}, z_3, \ldots, z_3\right),
\end{equation*}
where the within-group inequality of the second subgroup vanishes since all its members 
hold identical income shares. Then analogously to the case $n=3$ we could define the within-group residual $I^n\!\left(z_1,z_2,\frac{z_3}{n-2},\ldots,\frac{z_3}{n-2}\right)-  I^n\!\left(\frac{z_1+z_2}{2},\frac{z_1+z_2}{2}, z_3, \ldots, z_3\right)$, the between-group residual $I^n\!\left(z_1,z_2,\frac{z_3}{n-2},\ldots,\frac{z_3}{n-2}\right)- \frac{2}{n}I^2(z_1,z_2) $ and study their properties analogously to how we study their properties for $n=3$ in the next Section \ref{sec:g1g2properties}.

\section{Implications of classical decomposability notions}
\label{sec:g1g2properties}

Under decomposability  \eqref{eq:DCP} (\eqref{eq:DCI}), we have $g_1(z)=\frac{2}{3} I^2(z_1,z_2)$ ($g^{\mathcal{I}}_1(z)=I^2(z_1,z_2)$, which  implies several conditions that are useful for interpretation.
Some concern invariance and Schur-convexity, while others concern basic sign and identification properties.

\paragraph*{Properties of $g_1$ ($g_1^{\mathcal{I}}$)}

(1A) First,  under population-share-weighted  decomposability in \eqref{eq:DCP} (income-share-weighted decomposability in \eqref{eq:DCI}),  the within-group residual $g_1$ ($g_1^{\mathcal{I}}$) is invariant to redistributions that preserve relative income shares because for $z_1+z_2>0$, 
$$I^2(z_1,z_2)=I^2\left(\frac{z_1}{z_1+z_2},\frac{z_2}{z_1+z_2}\right) ,$$
On the simplex, this implies that under the respective decomposability, level sets of $g_1$ ($g_1^{\mathcal{I}}$) align with rays of constant $z_1/(z_1+z_2)$.

(1B), Second, by the definition of Schur-convexity for functions of two variables, under population-share-weighted (income-share-weighted) function $g_1$ ($g_1^{\mathcal{I}}$) should strictly increase with $|z_1-z_2|$ when $z_1+z_2$ remains constant.  

In addition to these properties, two basic interpretive conditions are implicit in the use of inequality decompositions. Under population-share-weighted (income-share-weighted) decomposability, 

(1C)  $g_1$ ($g_1^{\mathcal{I}}$) should be non-negative,

(1D) $g_1$ ($g_1^{\mathcal{I}}$) takes the value of zero  iff the corresponding source of heterogeneity is absent: that is, iff $z_1=z_2$. 

\paragraph*{Properties of $g_2$ ($g_2^{\mathcal{I}}$)}

(2A) Under population-share-weighted  decomposability in \eqref{eq:DCP} (income-share-weighted decomposability in \eqref{eq:DCI}) the between-group residual $g_2$ ($g_2^{\mathcal{I}}$) is  invariant along slices $z_3=\text{const}$: using $z_3=1-z_1-z_2$, under \eqref{eq:DCP} $g_2$ equals 
$$I^3\!\left(\frac{z_1+z_2}{2},\frac{z_1+z_2}{2},z_3\right)=I^3\!\left(\frac{1-z_3}{2},\frac{1-z_3}{2},z_3\right).$$
On the simplex, this implies that level sets of $g_2$ under \eqref{eq:DCP} ($g_2^{\mathcal{I}}$ under \eqref{eq:DCI}) align with slices of constant $z_3$.

(2B) Second,  Schur convexity of $I^3$ implies that under \eqref{eq:DCP} function $g_2$ (under \eqref{eq:DCI} function $g_2^{\mathcal{I}}$)
decreases in $z_3$ for $z_3 \in  [0, 1/3]$
(that is, when the average income in the subgroup of 2 is larger than the income in the subgroup of 1 but this gap is gradually reducing) and increases in $z_3$  for $z_3 \in [1/3, 1]$ (that is, when the average income in the subgroup of 2 is smaller than the income in the subgroup of 1 and this discrepancy is gradually widening). 

To show this mathematically, note that under \eqref{eq:DCP} we have $g_2(z)=I^3\!\left(\frac{1-z_3}{2},\frac{1-z_3}{2},z_3\right)$, and under \eqref{eq:DCI} we have $g_2^{\mathcal{I}}(z)=I^3\!\left(\frac{1-z_3}{2},\frac{1-z_3}{2},z_3\right)$. 
Denote 
$x(z) := \left( \frac{1-z}{2}, \frac{1-z}{2}, z \right)$, Recall that for vectors \(a,b \in \mathbb{R}^3\), writing
\(a^\downarrow, b^\downarrow\) for the components sorted in nonincreasing order,
\(a\) majorizes \(b\), denoted \(a \succ b\), if $\sum_{i=1}^k a_i^\downarrow \ge \sum_{i=1}^k b_i^\downarrow \quad (k=1,2)$,  $\sum_{i=1}^3 a_i^\downarrow = \sum_{i=1}^3 b_i^\downarrow.$
Strict Schur--convexity of \(I^3\) means that $a \succ b \implies I^3(a) > I^3(b)$. The sorted form of \(x(z)\) is 
\[
x(z_3)^\downarrow =
\begin{cases}
\left( \dfrac{1-z_3}{2}, \dfrac{1-z_3}{2}, z_3 \right), & 0 \le z_3 \le \dfrac13, \\[1ex]
\left( z_3, \dfrac{1-z_3}{2}, \dfrac{1-z_3}{2} \right), & \dfrac13 < z_3 \le 1 .
\end{cases}
\]
 First, take  \(0 \le {z}_3 < \widetilde{z}_3 \le \frac13\). Then $x(z) \succ x(\widetilde{z})$.
By strict Schur--convexity, $I^3(x(z_3)) > I^3(x(\widetilde{z}_3))$, so the value of $g_2(z)$   under \eqref{eq:DCP} (the value of $g_2^{\mathcal{I}}(z)$   under \eqref{eq:DCI}) must be decreasing in $z_3$ when  \(z_3 \in [0,\tfrac13]\). Now take \(\frac13 < z_3 < \widetilde{z}_3  \le 1\). Then $x(\widetilde{z}_3) \succ x(z_3)$. By strict Schur--convexity, $ I^3(x(\widetilde{z}_3)) > I^3(x(z_3))$, 
so \(g
_2(z)\) under \eqref{eq:DCP} ($g_2^{\mathcal{I}}(z)$   under \eqref{eq:DCI}) must be increasing in $z_3$ on \(z_3 \in [\tfrac13,1]\).

Just like in the case of $g_1$ ($g_1^{\mathcal{I}}$) we have two more interpretive conditions which are implicit in the use of inequality decompositions. Under population-share-weighted (income-share-weighted) decomposability, 

(2C)  $g_2$ ($g_2^{\mathcal{I}}$) should be non-negative.

(2D) $g_2$ ($g_2^{\mathcal{I}}$) should take the values of 0  iff the corresponding source of heterogeneity is absent: that is, iff the subgroup means are the same, or, equivalently,  $\frac{1-z_3}{2}=z_3$ or, equivalently, for any $z$  with $z_3=1/3$.

\paragraph*{Discussion} As we can see, these conditions cover a range of implications of decomposability and are based on various inequality measure properties. Condition (1A), (1B) and (2A), (2B) can perhaps be qualified as richer since they are examining the whole simplex. 

Conditions (1C), (1D) and (2C), (2D) may appear as more basic implications but nonetheless no less important. One might say that 
violations of either of these conditions indicate that the decomposition no longer has a natural interpretation. E.g. a violation of (2C) implies that subgroup mean differences contribute to a decrease of overall inequality under the chosen measure. 
Even when (1C) and (2C) hold, failure of (1D) or (2D), respectively, limits the interpretability of zero values.

\subsection{Graphical evaluation of decomposability implications}

Properties C and D  can be evaluated for any inequality measure by analyzing properties of $g_1$ and $g_2$ ($g_1^{\mathcal{I}}$ and $g_2^{\mathcal{I}}$) directly. Properties A and B are more conveniently evaluated graphically on the simplex $\Delta_2$.  To assess graphically the extent to which the functions $g_1$ and $g_2$ ($g_1^{\mathcal{I}}$ and $g_2^{\mathcal{I}}$) may deviate from properties A and B for a given inequality measure, I  propose the following collection of plots.

1. To assess (1A) and (1B) for population-share-weighted decomposability,  on the simplex $\Delta_2$, I plot the level sets (solid lines) 
\[
\{ z \in \Delta_2 : g_1(z) = c \}
\]
together with the rays (dashed lines) defined by
\[
\frac{z_1}{z_1+z_2} = \text{constant}.
\]
When (1A) holds, these two families of curves coincide exactly. 

For income-share-weighted decomposability, I plot the level sets 
\( z \in \Delta_2 : g_1^{\mathcal{I}} = c \}
\) of $g_1^{\mathcal{I}}$   
together with the aforementioned rays.

In addition, I superimpose a heat map in which larger values of $c$ are represented by progressively lighter colors. Under (1B), along any line in $\Delta_2$ with $z_3$ held constant, moving from the center point $z_1 = z_2 = \tfrac{1 - z_3}{2}$
toward the boundary of the simplex should result in a monotone transition to lighter colors only.

2. To assess (2A) and (2B) for population-share-weighted decomposability, on the simplex $\Delta_2$, I plot the level sets 
\[
\{ z \in \Delta_2 : g_2(z) = c \}.
\]
(solid lines). When (2A) holds, this family of curves should coincide with the family of slices $\{ z_3 = \text{constant} \}$. Whether this condition is satisfied can be readily assessed from the plot. 

For income-share-weighted decomposability, I plot the level sets   \[
\{ z \in \Delta_2 : g_2^{\mathcal{I}} = c \}
\] 
of $g_2^{\mathcal{I}}$ together with the slices constant in $z_3$.

With the heat map superimposed as described above, when (2B) holds, the following behavior should be observed: along any smooth curve in $\Delta_2$ that intersects each slice $\{ z_3 = \text{const} \}$ exactly once, the colors should vary monotonically from lighter to darker as $z_3$ increases up to $z_3 = \tfrac{1}{3}$, and then vary monotonically from darker to lighter as the curve crosses the slice $\{ z_3 = \tfrac{1}{3} \}$ and $z_3$ continues to increase from $\tfrac{1}{3}$ to $1$.

The simplex $\Delta_2$  in my graphical illustrations is  parameterized in
two dimensions using barycentric (equilateral–triangle) coordinates,
\[
(x,y)
=
\left(
z_2+\tfrac12 z_3,\;
\tfrac{\sqrt{3}}{2} z_3
\right),
\]
which maps $\Delta_2$ bijectively onto an equilateral triangle in $\mathbb{R}^2$. It is easy to see that $x$ changes from $0$ to 1, and $y$ changes from 0 to $\sqrt{3}/2$. The vertices of this equilateral triangle correspond to (i) $x=0, y=0$ (equivalently, $z_1=1$, $z_2=0$, $z_3=0$), (ii) $x=1, y=0$ (equivalently, $z_1=0$, $z_2=1$, $z_3=0$) and (iii) $x=1/2, y=\sqrt{3}/2$  (equivalently, $z_1=0$, $z_2=0$, $z_3=1$). 

Simplex $\Delta_2$ as the equilateral triangle in these barycentric coordinates appears in all the Figures \ref{fig:MLD_DCP}-\ref{fig:MLD_DCI} later in the paper. The base of the triangle always corresponds to cases $z_3=0$. The left panels in all figures also show the vertical line corresponding to $x=1/2$ which goes from the vertex $x=1/2, y=\sqrt{3}/2$ all the way to the base. Since 
$$x=\frac{1}{2} +\frac{z_2-z_1}{2},$$
such vertical line captures all the cases when $z_1=z_2$, that is, individual in the two-person subgroup have the same income.

\subsection{Reading the figures (summary)}

Each figure represents the income distribution $z=(z_1,z_2,z_3)$ as a point on the two-dimensional simplex $\Delta_ 2$, using barycentric coordinates described above. 

When evaluating property (1A) I plot the level sets of the within-group component $g_1$ ($g^{\mathcal{I}}_1$) as solid curves and also plot rays emanating from the vertex $x=1/2$, $y=\sqrt{3}/{2}$  that correspond to constant relative income shares within the two-person subgroup, that is, constant in $z_1/(z_1+z_2)$. These rays are depicted with the dashed lines.  Property (1A) means that the level sets of $g_1$ ($g^{\mathcal{I}}_1$)  coincide with the reference family of the dashed rays. 

When evaluating property (1B), we need to look at the color intensity which represents the magnitude of the component with  lighter colors indicating larger values.  Monotonic changes in color from darker to lighter from any location on the vertical line $x=1/2$ along an  orthogonal horizontal direction (either positive or negative) reflect the Schur-convexity property (1B). 

When evaluating property (2A)  I plot the level sets of the between-group residual $g_2$ ($g^{\mathcal{I}}_2$) as solid curves and horizontal slices (which are constancy relationships in $z_3$) as dashed lines. Property (2A) means that the level sets of $g_2$ ($g^{\mathcal{I}}_2$)  coincide with the reference family of the dashed horizontal slices. 

For $g_2$ plots, non-monotonic color patterns in the area $z_3<1/3$ (it should be from lighter to darker as $z_3$ increases) or in the area $z_3>1/3$ ((it should be from darker to lighter as $z_3$ increases)) along any curve on the simplex that crosses through each value of $z_3$ at most once  indicate violations of (2B), which also captures  Schur-convexity.

In this way, each figure simultaneously encodes invariance ((1A), (2A)) and   Schur-convexity (monotonicity (1B), (2B)) in our case $n=3$ in the decomposition.

\section{Population-share-weighted decomposability: graphical evaluation for common inequality measures}
\label{sec:population}

\subsection{Mean Log Deviation (GE(0))}

As established in the axiomatic literature, the Mean Log Deviation satisfies population--share--weighted decomposability, so this case serves as the benchmark for this decomposability properties and all (1A)-(1D) as well as (2A)-(2D) should be evidently satisfied. It is instructive to confirm these properties.   

 Using $z_3=1-z_1-z_2$, write functions $g_1$, $g_2$ as 
\begin{align*}
g_{1,MLD}(z) & = -\frac{1}{3}\ln \left( 4 \frac{z_1}{z_1+z_2}(1-\frac{z_1}{z_1+z_2})\right) \\
g_{2,MLD}(z) & =\frac{1}{3}\ln\!\left(\frac{4}{27}\right)
- \frac{1}{3}\ln z_3
- \frac{2}{3}\ln(1 - z_3)
\end{align*} 

\begin{figure}[ht]
    \centering
    \begin{minipage}{0.46\textwidth}
        \centering
        \includegraphics[width=\linewidth]{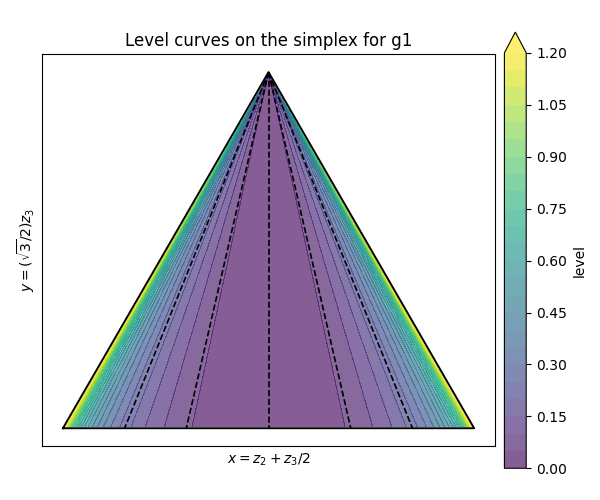}
    \end{minipage}
    \hfill
    \begin{minipage}{0.53\textwidth}
        \centering
        \includegraphics[width=\linewidth]{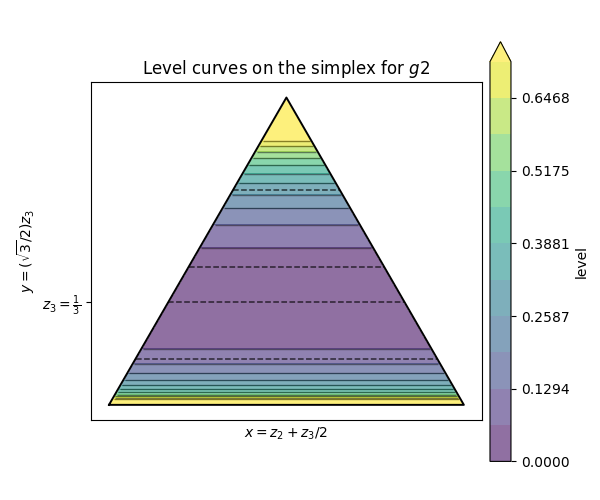}
    \end{minipage}
    \caption{\label{fig:MLD_DCP} Illustrations of decomposability \eqref{eq:DCP} for MLD . Left: Level curves for $g_{1,MLD}$ (solid lines) and rays constant at $\frac{z_1}{z_1+z_2}$ (dashed lines). Right: Level curves for $g_{2,MLD}$ (solid lines) and horizontal lines constant at $z_3$ (dashed lines).}
\end{figure}

Figure \ref{fig:MLD_DCP} shows that the Mean Log Deviation satisfies both geometric restrictions. 

The graph on the left-hand side of Figure \ref{fig:MLD_DCP} shows that level sets of $g_1$ align exactly with rays of constant $z_1/(z_1+z_2)$ confirming (1A). It also visually confirms (1B). If we start from any point on the vertical line $x=\frac{1}{2}$ (equivalently, $z_1=z_2$) in the middle and move from that point   in either of the two horizontal directions (thus, keeping $z_3=1-(z_1+z_2)$ fixed) towards the simplex boundary the colour changes monotonically from darker to lighter. 

The graph on the right-hand side of Figure \ref{fig:MLD_DCP} shows that level sets of  $g_2$ align with horizontal slices corresponding to constant $z_3$, thus, confirming (2A). To illustrate (2B), notice the location  of the tick corresponding to $z_3=1/3$ on the vertical axis. Visualize a monotonic curve on the simplex that crosses each $\{z_3=constant\}$ at most once. If we move from the bottom of the simplex, we will see colours on this curve getting gradually darker until the line $\{z_3=1/3\}$ and then getting gradually  lighter once this line is crossed.  This confirms (2B). 

Properties (1C), (1D) and (2C). (2D) can be easily verified analytically.

Thus, to sum up, this case serves as a benchmark illustrating how exact population-share-weighted decomposability manifests geometrically.

\subsection{Gini} 
\label{sec:GiniDCP} 
The Gini coefficient is known to lack population-share-weighted  decomposability. It will  therefore come as no surprise that some of the properties of $g_1$ and $g_2$ will be violated.\footnote{Gini's nature is  rank-based and  it introduces an ``overlap'' or ``interaction'' term in subgroup decompositions, as noted in early critiques like \cite{Pyatt1976} and extended in other  works later.} The simplex framework clarifies the source of this failure. For Gini, 
\begin{align*}
g_{1,Gini}(z) & =\frac{1}{3}\left( |z_1-z_2|+|z_1-z_3|+|z_2-z_3| \right)-\frac{2}{3} |(z_1+z_2)/2-z_3|\\
& = \frac{1}{3}\left( |z_1-z_2|+|z_1-z_3|+|z_2-z_3| \right)-\frac{1}{3} |3z_3-1|\\
& \left\{ 
\begin{array}{l}
\frac{1}{3}|z_1-z_2|, \quad \text{if } z_3\geq \max\{z_1,z_2\} \text{ or } z_3\leq \min\{z_1,z_2\} \\
\frac{2}{3}|z_1-z_2| -\frac{1}{3} |3z_3-1|, , \quad \text{if } z_3\in (\min\{z_1,z_2\}, \max\{z_1,z_2\} )
\end{array}
\right. \\ 
g_{2,Gini}(z) & =\frac{1}{3}\left( |z_1-z_2|+|z_1-z_3|+|z_2-z_3| \right) -\frac{1}{3}\left( |z_1-z_2| \right)/(z_1+z_2)
\end{align*} 

\begin{figure}[ht]
    \centering
    \begin{minipage}{0.47\textwidth}
        \centering
        \includegraphics[width=\linewidth]{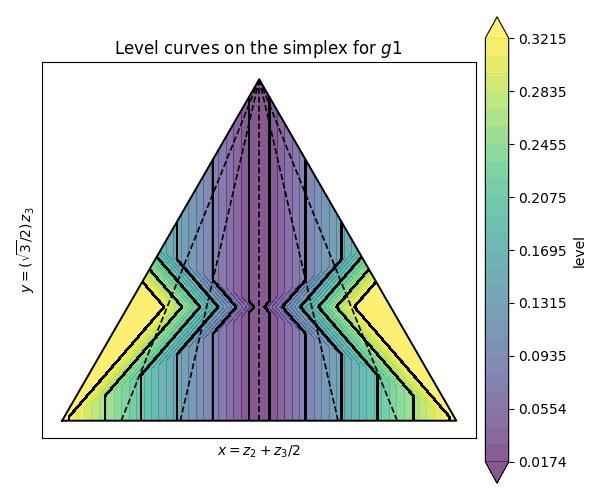}
    \end{minipage}
    \hfill
    \begin{minipage}{0.52\textwidth}
        \centering
        \includegraphics[width=\linewidth]{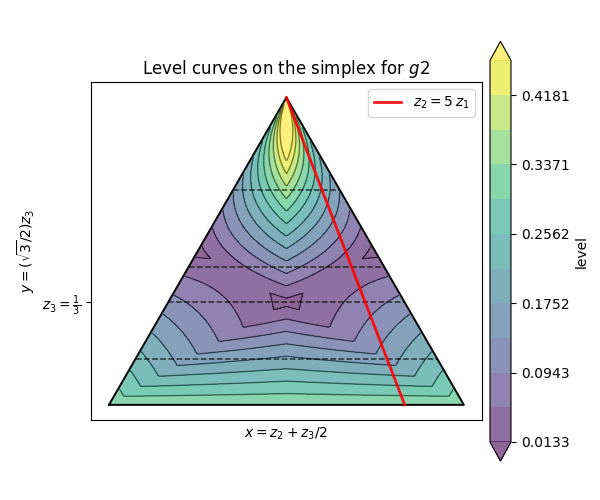}
    \end{minipage}
    \caption{\label{fig:GiniDCP} Illustrations for Gini. Left: Level curves for $g_{1,Gini}$ and rays constant at $\frac{z_1}{z_1+z_2}$. Right: Level curves for $g_{2,Gini}$.}
\end{figure}

Figure \ref{fig:GiniDCP} shows that to what extent the Gini coefficient violates  geometric restrictions. 

\paragraph*{Properties (1A)-(1D)}
The plot on the left-hand side of Figure \ref{fig:GiniDCP} clearly shows the extent of violation of (1A): level curves for Gini are distinct from rays constant at $z_1/(z_1+z_2)$. Thus, redistribution within $z$ that leaves $z_1/(z_1+z_2)$ invariant affects the within-group inequality. In other words,  the within-group inequality is contaminated by the third individual's position. This violation of (1A) occurs throughout the whole simplex and is not limited to a particular area of the simplex. Violations in the case of rank overlap ($\min\{z_1,z_2\}<z_3<\max\{z_1,z_2\}$) appear worse than in cases when $z_3$ in the solo group either fully dominates or is fully dominated by $z_1$, $z_2$ in the two-person group. 

The same  plot also confirms that (1B) holds -- taking any point on the vertical line $x=1/2$ ($\iff z_1=z_2$), move from that point in either horizontal direction towards the simplex boundary and notice colors getting monotonically lighter.

At the same time, the Gini coefficient satisfies non-negativity in (1C) because by the triangle inequality, $3g_1(z) \geq |z_1-z_2|\geq 0$, and satisfies (1D) since the same inequality guarantees that $g_1(z)=0$ implies $z_1=z_2$. Thus, the zero of the within-group residual is informative and does indeed indicate that the incomes within  the subgroup are the same. 

\paragraph*{Properties (2A)-(2D)} 
The plot on the right-hand side of Figure \ref{fig:GiniDCP}  shows extent of the violation of (2A): level curves for Gini are distinct from horizontal lines corresponding to constant  $z_3$. 

That plot also illustrates that (2B)  does not hold.  It displays the tick on the vertical axis  corresponding to $z_3=1/3$ and gives an example of the curve $z_2=5z_1$ (red) which is monotonic on the simplex.  If we look at the movement along this curve starting from $z_3=1/3$ and gradually increasing $z_3$, we see that first, the colors change from lighter to darker and then gradually to lighter colors. Under (2B), in that area the change in colors would have been  monotonic from darker to lighter. The violation presented here is in the area $z_3>1/3$ (when the income in the one-person subgroup dominates the average income in the two-person subgroup) but it can also be constructed for $z_3<1/3$ (when the two-person subgroup average dominates the singleton). 

Gini satisfies (2C) but not (2D):
for example, $z^a=\Bigl(\frac12,0,\frac12\Bigr)$ 
 and 
$z^b=\Bigl(0,\frac12,\frac12\Bigr)$ yield \(g_2(z)=0\). The violation can also be seen from the right-hand side of Figure \ref{fig:GiniDCP} where there  are two dark areas around $z^a$ and $z^b$. Thus, the between-group residual may equal zero at income configurations that do not equate subgroup means across different subgroups. In practice, this means that a zero between-group residual
does not necessarily indicate an absence of group differences. At the same time, the heat map on the plot in the right-hand side of Figure \eqref{fig:GiniDCP} shows another aspect of (2D) violation by Gini:  that there are many $z$ on the horizontal  line $z_3=1/3$ that don't give the value 0 to the between residual. In practice, this means that a positive between-group residual 
does not necessarily indicate a presence of group differences.

Thus, our analysis illustrates geometrically how exactly population-share weighting, which  in a way  treats each individual as contributing equally to inequality, independently of income rank,  clashes with the defining rank-based feature of the Gini coefficient.

\subsection{Coefficient of variation} 
\begin{figure}[ht]
    \centering
    \begin{minipage}{0.47\textwidth}
        \centering
        \includegraphics[width=\linewidth]{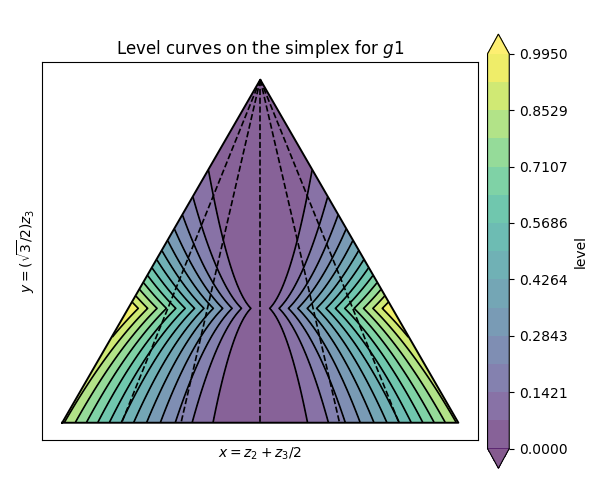}
    \end{minipage}
    \hfill
    \begin{minipage}{0.52\textwidth}
        \centering
        \includegraphics[width=\linewidth]{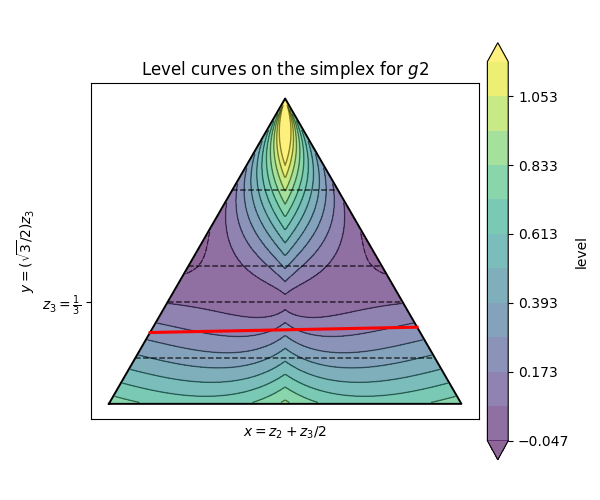}
    \end{minipage}
    \caption{\label{fig:CV_DCP} Illustrations for CV. Left: Level curves for $g_{1,CV}$ and rays constant at $\frac{z_1}{z_1+z_2}$. Right: Level curves for $g_{2,CV}$.}
\end{figure}

For the coefficient of variation defined as 
\(
I^n(y_1,\ldots, y_n)=\frac{\sqrt{\frac{1}{n-1}\sum_{i=1}^n (y_i-\bar y)^2}}{\bar y}, 
\) our functions take the form 
\begin{align*}
g_{1,CV}(z) & =\sqrt{\frac{1}{2}\sum_{i=1}^3 (3z_i-1)^2} - \frac{\sqrt{3}}{2} |3z_3-1|, \\
g_{2,CV}(z) & =\sqrt{\frac{1}{2}\sum_{i=1}^3 (3z_i-1)^2}  -\frac{2\sqrt{2}}{3}|2 \frac{z_1}{z_1+z_2}-1|.
\end{align*} 
Figure \ref{fig:CV_DCP} illustrates the degrees of violations of decomposability for the coefficient of variation.

\paragraph*{Properties (1A)-(1D)}
The left-hand side plot in Figure \ref{fig:CV_DCP} shows how the level curves of $g_{1,CV}$ are  distinct from rays constant at $z_1/(z_1+z_2)$. Thus, (1A) does not hold. The violations are particularly drastic when $z_3>1/3$ -- that is, when the income in solo groups dominates incomes in the two-person group. 

That plot also makes it clear that (1B) holds -- thus, our $g_1$ has monotonicity properties with the growing gap between $z_1$ and $z_2$ holding $z_1+z_2=1-z_3$ constant. 

(1C) holds as for any fixed $z_3$ the function $\frac{1}{2}\sum_{i=1}^3 (3z_i-1)^2$ is minimized when $z_1=z_2=\frac{1-z_3}{2}$. Hence, for a fixed $z_3$ the minimum value of  $\sqrt{\frac{1}{2}\sum_{i=1}^3 (3z_i-1)^2}$ is $\frac{\sqrt{3}}{2} |3z_3-1|$. Consequently, $g_{1,CV}(z) \geq 0$. This discussion also shows that $g_1(z)=0$ iff  $z_1=z_2$. Hence, (1D) holds too. 

\paragraph*{Properties (2A)-(2D)}
The properties of the between-group residual  $g_{2,CV}$  are much more severely violated compared to the within-group residual  $g_{1,CV}$.  

(2A) does not hold as from the right-hand side plot of Figure \ref{fig:CV_DCP} it is obvious that level of $g_{2,CV}$ are not horizontal. That plot also makes it clear that (2B) does not hold either. It shows the level (dashed horizontal line) corresponding to $z_3=1/3$ and gives an example of the curve  (red) which is monotonic on the simplex.  If we look at the movement along this curve  increasing in $z_3$ (it is always below $1/3$), we see that first, the colors change from lighter to darker and then at some point move back from a darker to a lighter color.   We can also construct curves with clear violations in the area $z_3>1/3$. Thus, the between-group residual may decrease (increase) following redistributions that increase (decrease) differences in subgroup means.

(2C) does not hold as $g_{2,CV}$ can take negative values -- e.g. for $z=(0,1/2,1/2)$ or $z=(1/2,0,1/2)$. When the 
between–group residual is negative, one might conclude that it no longer admits a natural interpretation as a contribution to inequality.

(2D) does not  hold in two different ways. First, the zeros of $g_{2,CV}$ may happen at  points with $z_3\neq 1/3$.\footnote{ Namely, $g_{2,CV}(z)=0$ if $z=(1/3,1/3,1/3)$ or when  $$z_1 = \frac{s}{2}\pm\frac12
\sqrt{\frac{243\,s^2\left(s-\tfrac23\right)^2}{32-81s^2}}, \quad z_2 = \frac{s}{2}\mp\frac12
\sqrt{\frac{243\,s^2\left(s-\tfrac23\right)^2}{32-81s^2}},$$ $z_3 = 1-s$, and $s\in\left[\frac{9-\sqrt5}{18},\frac{9+\sqrt5}{18}\right],\quad s\neq\frac23$. }
Thus, the zero values of this between-group residual will not necessarily imply the equality of subgroup means. 

On the other hand, as implied by the analysis  above the points on the horizontal line $z_3=1/3$ different from $(1/3,1/3,1/3)$ will not deliver the value of 0.  Thus, most of the cases of subgroup means equality will be assigned  a positive between-group residual. 

Thus, the coefficient of variation appears ill-suited for the population-share-weighted subgroup decomposition exercise. In addition to non-monotonicity in subgroup mean differences, which is shared by other non-decomposable measures, 
the between–group residual can be negative. 
From an applied perspective, this highlights a limitation of variance-based measures for subgroup analysis.\footnote{The variance itself is naturally decomposed into the within/between components but applying nonlinear transformation (taking square root) and dividing it by the mean to ensure scale-invariance introduces nonlinearities that lead to particular poor properties of the between-group residual.}

\subsection{Theil index (GE(1))} As is well known, while the Theil index satisfies income-share-weighted decomposability, 
it fails population-share-weighted decomposability. This  will be reflected in geometric properties of functions $g_1$ and $g_2$.  First of all, for Theil, 
\begin{align*}
g_{1,Theil}(z) & =z_1 \ln\left(3 z_1\right) + z_2\ln\left(3 z_2\right) - \frac{z_1+z_2}{2} \ln\left(\frac{3}{2} (z_1+z_2)\right) - \frac{z_1+z_2}{2}\ln\left(\frac{3}{2} (z_1+z_2)\right) \\
&= (z_1+z_2) KL\left(\frac{z_1}{z_1+z_2}, 1- \frac{z_1}{z_1+z_2} \mid (\frac{1}{2},\frac{1}{2})\right) \\
g_{2,Theil}(z) & =z_1 \ln\left(3 z_1\right) + z_2\ln\left(3 z_2\right) + z_3\ln\left(3 z_3\right)   -\frac{2z_1}{3(z_1+z_2)} \ln\left(2 \frac{z_1}{z_1+z_2}\right) - \frac{2z_2}{3(z_1+z_2)}\ln\left(2 \frac{z_2}{z_1+z_2}\right)
\end{align*} 
The expression for $g_{1,Theil}$ uses Kullback–Leibler (KL) divergence (relative entropy) between two discrete probability distributions: if 
$p = (p_1,\dots,p_n)$, $q = (q_1,\dots,q_n)$
are probability distributions on a finite set, then KL  divergence from \(q\) to \(p\) is defined as
\(
KL(p \,\|\, q)
\;=\;
\sum_{i=1}^{n} p_i
\ln\!\left(\frac{p_i}{q_i}\right),
\)

\begin{figure}[ht]
    \centering
    \begin{minipage}{0.47\textwidth}
        \centering
        \includegraphics[width=\linewidth]{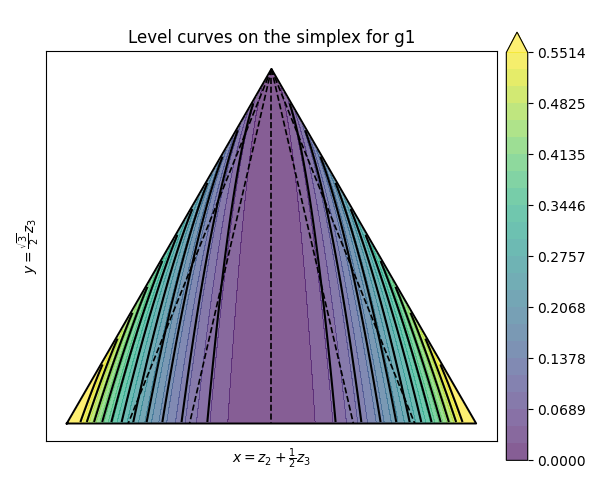}
    \end{minipage}
    \hfill
    \begin{minipage}{0.52\textwidth}
        \centering
        \includegraphics[width=\linewidth]{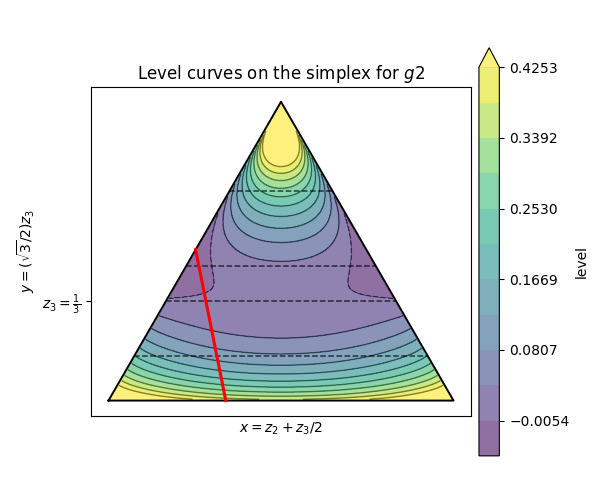}
    \end{minipage}
    \caption{\label{fig:Theil_DCP} Illustrations for Theil (GE(1)). Left: Level curves for $g_{1,Theil}$ and rays constant at $\frac{z_1}{z_1+z_2}$. Right: Level curves for $g_{2,Theil}$.}
\end{figure}

\paragraph*{Properties (1A)-(1D)}
Geometrically, as can be seen from the left-hand side plot in Figure \ref{fig:Theil_DCP}, level sets of $g_{1,Theil}$ exhibit approximate but not exact invariance with the rays constant at $z_1/(z_1+z_2)$. Thus, even though (1A) fails, it does so in a structured way. This is a much milder violation compared to those we have seen in Gini and CV for this property. 

As confirmed by the same plot, (1B) holds. The mechanics in  which it can be seen is  exactly the same way as for the previous indices we have discussed. 

 (1C) evidently holds from the KL representation of $g_{1,Theil}$. 
 
 The minimum value 0 is attained when either $z_1+z_2=0$ (that is, $z_1=z_2=0$) or $z_1+z_2 \neq 0$ and $z_1=z_2$. These situations describe a generic case $z_1=z_2$. Thus, (1D) holds as well.

\paragraph*{Properties (2A)-(2D)} 
For properties of $g_{2,Theil}$, (2A) obviously  does not hold -- from the right panel of Figure \ref{fig:Theil_DCP} it is obvious how level curves of $g_{2,Theil}$ are different from horizontal slices. 

The same plot visually confirms that (2B) does not hold. It shows the level (horizontal line) corresponding to $z_3=1/3$ and gives an example of the curve  (red) which is monotonic on the simplex with the following property.   If we look at the movement along this curve  increasing in $z_3$ from $z_3=1/3$, we see that first the colours change from lighter to darker and then at some point change from darker to lighter. Even though in this illustration violations come from the region $z_3 > 1/3$, we could construct monotonicity violations in the region $z_3<1/3$ as well.

(2C) does not hold and $g_{2,Theil}$ can take negative values -- e.g. for $z=(0,1/2,1/2)$ or $z=(1/2,0,1/2)$. 

(2D) does not hold fully. First, what holds is that the entire segment $z_3=\frac{1}{3}$ (equivalently $z_1+z_2=\frac{2}{3}$) gives the zero value to $g_{2,Theil}$. However, there are points $z$ outside this segment that also give zero values to $g_{2,Theil}$. Namely, for $z_3=\frac{2}{3}$,  
 the two points $\left(\frac{1}{3},\,0,\,\frac{2}{3}\right)$ and $\left(0,\,\frac{1}{3},\,\frac{2}{3}\right)$ also result in the zero values for  $g_{2,Theil}$. Moreover, for each $z_3 \in (\frac{1}{3},\frac{2}{3})$ (equivalently,  $z_1+z_2 \in (\frac{1}{3},\frac{2}{3})$), we can find $z_1+z_2=1-z_3$ such that   
\[
\frac{z_1}{z_1+z_2}\ln \frac{z_1}{z_1+z_2}+\frac{z_2}{z_1+z_2}\ln \frac{z_2}{z_1+z_2}
=
-\frac{A(z_1+z_2)}{\,z_1+z_2-\frac{2}{3}\,},
\]
with
\(
A(s)=\ln 3-\frac{2}{3}\ln 2+s\ln s+(1-s)\ln(1-s).
\)
Such points will also give zero values to $g_{2,Theil}$. E.g., for $z_3=0.5$  we can take one of $z_1$, $z_2$ to be $0.05340609$ and the other one to be $0.44659391$, etc. In a nutshell, the zero value of the between component will not necessarily imply equality of subgroup means.

\subsection{Discussion}

The analysis clarifies that population-share-weighted decomposability is not a binary property.
Inequality measures differ in the structure of their departures from this decomposability property. 

For Gini, failure arises from the dependence on global rank distances.
For CV, failure arises from nonlinear transformation of variance and the sensitivity to extreme outcomes.
For the Theil index, failure is driven by the choice of the weighting scheme.
These findings are consistent with the existing literature but add precision by localising failures on the simplex.

We have seen that across all three non-decomposable measures  in the population-share-weighted sense (that is, Gini, CV, Theil) the within-group residual $g_1$ preserved the Schur-convexity as stipulated in (1B) and the   worst performance across all these measures occurs for the between-group residual $g_2$. 

With the graphical diagnostics revealing  that violations differ in structure and magnitude, it would be natural to try and define measures of ``distance from decomposability,'' for example by quantifying deviations of level sets from the reference rays or slices. Such metrics are beyond the scope of this  paper, but the geometric framework developed here provides a natural starting point for formalising degrees of decomposability. 

The summary of violations is given in Table \ref{table:DCP}.

\begin{table}[H]
\centering
\begin{threeparttable}
\caption{Population-share-weighted decomposability (DCP): summary of property failures}
\label{table:DCP}
\begin{tabular}{lcccccccc}
\toprule
Measure & (1A) & (1B) & (1C) & (1D) & (2A) & (2B) & (2C) & (2D) \\
\midrule
MLD   & \checkmark & \checkmark & \checkmark & \checkmark & \checkmark & \checkmark & \checkmark & \checkmark \\
Gini  & \texttimes & \checkmark & \checkmark & \checkmark & \texttimes & \texttimes & \checkmark & Z$\Rightarrow$E fails; E$\Rightarrow$Z fails \\
CV    & \texttimes & \checkmark & \checkmark & \checkmark & \texttimes & \texttimes & \texttimes & Z$\Rightarrow$E fails; E$\Rightarrow$Z fails \\
Theil & $\approx$  & \checkmark & \checkmark & \checkmark & \texttimes & \texttimes & \texttimes & Z$\Rightarrow$E fails \quad \quad \quad \quad \; \; \\
\bottomrule
\end{tabular}
\begin{tablenotes}
\footnotesize
\item \textit{Notes:} 
Z$\Rightarrow$E means ``zero between-group component implies equal subgroup means.'' 
E$\Rightarrow$Z means ``equal subgroup means imply zero between-group component.'' 
$\approx$ denotes structured but inexact invariance. 
\checkmark\ indicates the property holds; \texttimes\ indicates failure.
\end{tablenotes}
\end{threeparttable}
\end{table}

\section{Income-share-weighted decomposability: graphical evaluation for common inequality measures}
\label{sec:income}

For income-share-weighted decomposability, we have to look at the properties of $g^{\mathcal{I}}_{1}$ and $g^{\mathcal{I}}_{2}$. 

Since it is well known that for this notion of decomposability, the Theil index (GE(1)) is the only one that satisfies this property, this is our benchmark case which we consider first, then Gini, then CV and, finally, Mean Log Deviation.  

\subsection{Theil index} 
\begin{figure}[ht]
    \centering
    \begin{minipage}{0.47\textwidth}
        \centering
        \includegraphics[width=\linewidth]{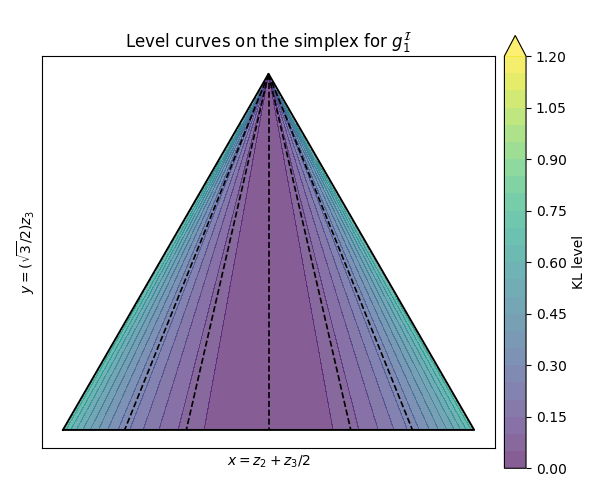}
    \end{minipage}
    \hfill
    \begin{minipage}{0.52\textwidth}
        \centering
        \includegraphics[width=\linewidth]{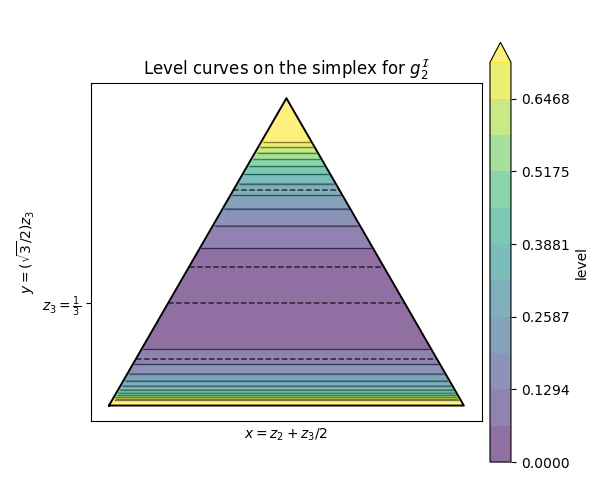}
    \end{minipage}
    \caption{\label{fig:TheilDCI} Illustrations for Theil. Left: Level curves for $g^{\mathcal{I}}_{1,Theil}$ and rays constant at $\frac{z_1}{z_1+z_2}$. Right: Level curves for $g^{\mathcal{I}}_{2,Theil}$.}
\end{figure}

For this index, 
\begin{align*}
g^{\mathcal{I}}_{1,Theil}(z) &  =KL\left(\frac{z_1}{z_1+z_2}, \frac{z_2}{z_1+z_2} \mid (\frac{1}{2},\frac{1}{2})\right) \\
g^{\mathcal{I}}_{2,Theil}(z) & =z_1 \ln\left(3 z_1\right) + z_2\ln\left(3 z_2\right) + z_3\ln\left(3 z_3\right)   -z_1 \ln\left(2 \frac{z_1}{z_1+z_2}\right) - z_2\ln\left(2 \frac{z_2}{z_1+z_2}\right) \\
&=(z_1+z_2)\ln (\frac{3}{2}(z_1+z_2)) +z_3\ln\left(3 z_3\right) =(1-z_3)\ln (\frac{3}{2}(1-z_3))  +z_3\ln\left(3 z_3\right)
\end{align*} 
Figure \ref{fig:TheilDCI} shows that the Theil index satisfies  geometric restrictions (1A), (1B) and (2A), (2B) for $g^{\mathcal{I}}_{1,Theil}$ and $g^{\mathcal{I}}_{2,Theil}$, respectively.  

Properties (1C), (1D) and (2C), (2D) can be checked theoretically to hold. 
\subsection{Gini} 
We have 
\begin{align*}
g^{\mathcal{I}}_{1,Gini}(z) & = \frac{\frac{1}{3}\left( |z_1-z_2|+|z_1-z_3|+|z_2-z_3| \right)-\frac{1}{3} |3z_3-1|}{z_1+z_2}\\
& =\left\{ 
\begin{array}{l}
\frac{1}{3}|\frac{z_1}{z_1+z_2}-\frac{z_2}{z_1+z_2}|, \quad \text{if } z_3\geq \max\{z_1,z_2\} \text{ or } z_3\leq \min\{z_1,z_2\} \\
\frac{2}{3}|\frac{z_1}{z_1+z_2}-\frac{z_2}{z_1+z_2}| -\frac{1}{3} \frac{|3z_3-1|}{z_1+z_2} , \quad \text{if } z_3\in (\min\{z_1,z_2\}, \max\{z_1,z_2\} ), 
\end{array}
\right. \\ 
g^{\mathcal{I}}_{2,Gini}(z) & =\frac{1}{3}\left( |z_1-z_2|+|z_1-z_3|+|z_2-z_3| \right) -\frac{1}{2} |z_1-z_2|.  
\end{align*} 

\begin{figure}[ht]
    \centering
    \begin{minipage}{0.47\textwidth}
        \centering
        \includegraphics[width=\linewidth]{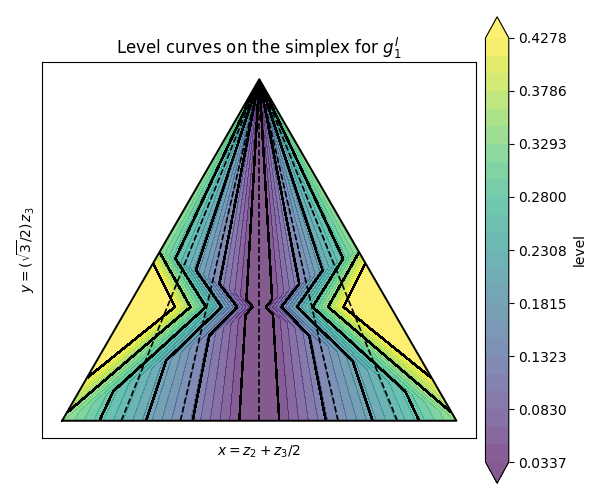}
    \end{minipage}
    \hfill
    \begin{minipage}{0.52\textwidth}
        \centering
        \includegraphics[width=\linewidth]{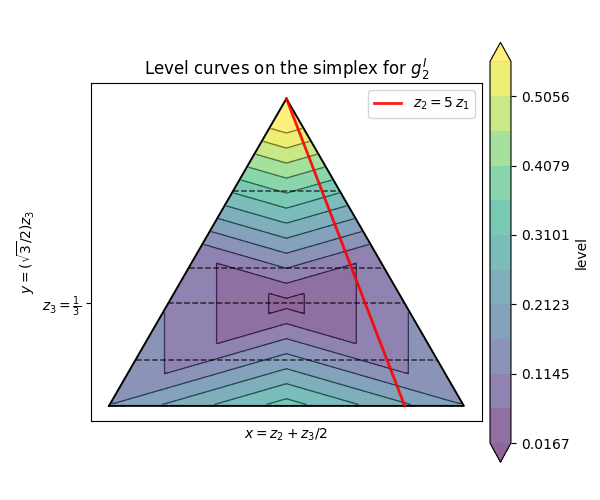}
    \end{minipage}
    \caption{\label{fig:GiniDCI} Illustrations for Gini. Left: Level curves for $g^{\mathcal{I}}_{1,Gini}$ and rays constant at $\frac{z_1}{z_1+z_2}$. Right: Level curves for $g^{\mathcal{I}}_{2,Gini}$.}
\end{figure}

Figure \ref{fig:GiniDCI} shows to what extent the Gini coefficient violates  geometric restrictions. 

\paragraph*{Properties (1A)-(1D)} From the left panel of Figure \ref{fig:GiniDCI} it is obvious that (1B) holds. Property (1A) holds only in the subset of the simplex where $z_3\geq\max\{z_1,z_2\}$ or $z_3\leq\min\{z_1,z_2\}$. Thus, $g^{\mathcal{I}}_{1,Gini}$ satisfies the key invariance condition (1A) on large regions of the simplex -- specifically, whenever the singleton income either dominates or is dominated by both members of the two-person group. In these regions, rank orderings are stable, and the Gini behaves “as if” it were decomposable. The remaining violations arise precisely where rank reversals occur, i.e., when the singleton income lies between the two subgroup incomes ($\min\{z_1,z_2\}<z_3<\max\{z_1,z_2\}$). This localisation of failure is much sharper than under population-share weighting, where violations occur everywhere.

(1C) holds: using the fact that $1=z_1+z_2+z_3$ in the $|3z_3-1|$ term, we get that 
\begin{align*} g^{\mathcal{I}}_{1,Gini}(z) &= \frac{2(\max\{z_1,z_2,z_3\}- \min\{z_1,z_2,z_3\})-|3z_3-1|}{3(z_1+z_2)} \\
& = \frac{2(\max\{z_1,z_2,z_3\}- \min\{z_1,z_2,z_3\})-2|z_3-\frac{z_1+z_2}{2}|}{3(z_1+z_2)} \geq 0
\end{align*}
as 
$\max\{z_1,z_2,z_3\}- \min\{z_1,z_2,z_3\} \geq \left|z_3-\frac{z_1+z_2}{2}\right|.$ 

(1D) holds because in light of the analysis above  $g^{\mathcal{I}}_{1,Gini}(z)$ will attain the value of 0 iff $\max\{z_1,z_2,z_3\}- \min\{z_1,z_2,z_3\} =  \left|z_3-\frac{z_1+z_2}{2}\right|$, which happens iff $z_1=z_2$. 

\paragraph*{Properties (2A)-(2D)} 

The extent of violations of (2A) and (2B) can be seen on the right-hand side plot in Figure \ref{fig:GiniDCI} in a way completely analogous to the earlier discussion of $g_{2,Gini}$ properties in Section \ref{sec:GiniDCP}. In  particular, the red curve illustrates violations of (2B). This particular displayed curve illustrates violations in the area $z_3 > 1/3$ but analogous violations can be constructed for $z_3 \leq 1/3$ as well.   

(2C) holds. Indeed, 
$
g^{\mathcal{I}}_{2,Gini}(z)
=\frac13\bigl(|z_1-z_3|+|z_2-z_3|\bigr)
-\frac16|z_1-z_2|.$ 
By the triangle inequality, $|z_1-z_2|
\le |z_1-z_3|+|z_2-z_3|$. Hence,
$g^{\mathcal{I}}_{2,Gini}(z)
\ge \frac13|z_1-z_2|-\frac16|z_1-z_2| 
=\frac16|z_1-z_2|
\ge 0.$
Therefore, $g^{\mathcal{I}}_{2,Gini}(z)$ on the simplex is non-negative. 

(2D) does not hold, however. If $g^{\mathcal{I}}_{2,Gini}(z)=0$, then from the bound above we must have simultaneously
$|z_1-z_2|=0$ and also we must have equality in the triangle inequality used, which would force $|z_1-z_3|+|z_2-z_3|=0$, hence $z_1=z_2=z_3$.  In other words, $(\frac{1}{3}, \frac{1}{3},\frac{1}{3})$ is the only point in the zero set of $g^{\mathcal{I}}_{2,Gini}$. Therefore,  most of the cases of equal subgroup means will result in a strictly positive between component.

Thus, the behavior of the within-group residual $g^{\mathcal{I}}_{1,Gini}$ shows that income-share decomposability may be considered as better aligned with the logic of the Gini. In the simplex representation, we see a markedly different pattern of violations than in the population-share-weighted case: $g_1^{\mathcal{I}}$ satisfies the  monotonicity and zero conditions everywhere and satisfies the key invariance property except in regions where rank reversals occur (that is, when the singleton income lies between the two subgroup incomes). In other words, departures from decomposability become  economically interpretable, rather than generic. The between-group residual $g^{\mathcal{I}}_{2,Gini}$ under income-share weighting also behaves more coherently. Property (2D) still fails but only in one way (equal subgroup means do not generally imply a zero between component).

\subsection{Coefficient of variation}
We have 
\begin{align*}g^{\mathcal{I}}_{1,CV} & =\frac{\sqrt{\frac{1}{2}\sum_{i=1}^3 (3z_i-1)^2}- 3\sqrt{3}/2 |z_3-1/3|}{z_1+z_2} \\ 
&= \frac{3}{2\,(1-z_3)}
\left(
\sqrt{(z_1-z_2)^2+3\left(z_3-\frac13\right)^2}
-\sqrt{3}\left|z_3-\frac13\right|
\right) \\ 
g^{\mathcal{I}}_{2,CV}&=\sqrt{\frac{1}{2}\sum_{i=1}^3 (3z_i-1)^2}- (z_1+z_2)\sqrt{2}|2 \frac{z_1}{z_1+z_2}-1|
\end{align*}

\begin{figure}[ht]
    \centering
    \begin{minipage}{0.47\textwidth}
        \centering
        \includegraphics[width=\linewidth]{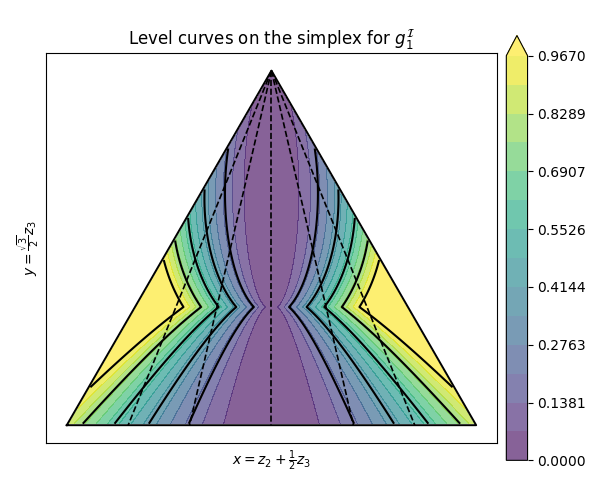}
    \end{minipage}
    \hfill
    \begin{minipage}{0.52\textwidth}
        \centering
        \includegraphics[width=\linewidth]{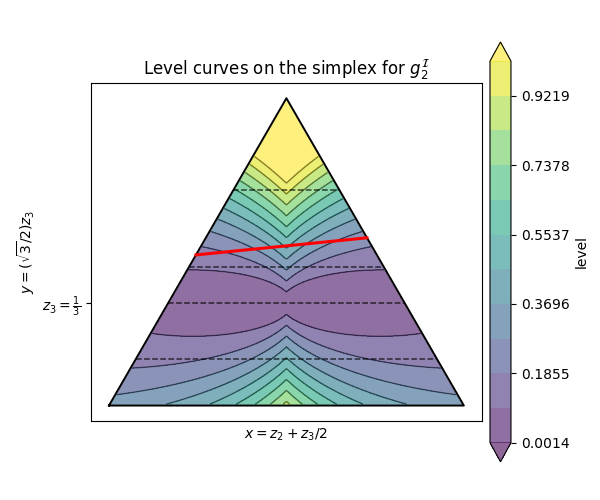}
    \end{minipage}
    \caption{\label{fig:CV_DCI} Illustrations for Coefficient of Variation. Left: Level curves for $g^{\mathcal{I}}_{1,CV}$ and rays constant at $\frac{z_1}{z_1+z_2}$. Right: Level curves for $g^{\mathcal{I}}_{2,CV}$.}
\end{figure}

Figure \ref{fig:CV_DCI} shows to what extent the coefficient of variation violates  geometric restrictions. 

\paragraph*{Properties (1A)-(1D)}
The left panel of Figure \ref{fig:CV_DCI} shows the extent in which (1A) is violated. It confirms that (1B) holds.  

(1C) holds. Note that $(3z_1-1)^2+(3z_2-1)^2\ge \frac{(3z_1-1+3z_2-1)^2}{2}=\frac{(3z_3-1)^2}{2}$, hence, $(3z_1-1)^2+(3z_2-1)^2 + (3z_3-1)^2\ge \frac{3(3z_3-1)^2}{2}$, and, therefore, 
\[
\sqrt{\frac{(3z_1-1)^2+(3z_2-1)^2 + (3z_3-1)^2}{2}}
\ge
\frac{\sqrt3}{2}\lvert 3z_3-1\rvert,
\]
implying that $g^{\mathcal{I}}_{1,CV}(z) $ is non-negative. 

(1D) holds as well. Based on our analysis for (1C) we can conclude that 
$g^{\mathcal{I}}_{1,CV}(z)=0$ holds if and only if $(3z_1-1)^2+(3z_2-1)^2= \frac{(3z_1-1+3z_2-1)^2}{2}$, i.e. $3z_1-1=3z_2-1=-\frac{3z_3-1}{2},$ which is equivalent to $z_1=z_2=\frac{1-z_3}{2}$.

\paragraph*{Properties (2A)-(2D)} 

The right panel of Figure \ref{fig:CV_DCI} shows the degree of violations of (2A) and the pattern of distinction between level curves  from the horizontal lines.  (2B) is violated as well as exemplified by the red curve which is monotonic in the simplex  but that change in the heat map on that curve is not monotonic. This curve illustrates violations of (2B) in the region $z_3>1/3 \iff z_3 >\frac{z_1+z_2}{2}$, but a similar  illustration can be given for $z_3<1/3$.  

(2C) holds. First of all, notice that $z_1+z_2+z_3=1$ implies 
$$(3z_1-1)^2 +(3z_2-1)^2 =\frac{1}{2}(3z_3-1)^2+\frac{1}{2}(3z_1-1-(3z_2-1))^2,$$
hence, 
$\sqrt{\frac{(3z_1-1)^2+(3z_2-1)^2 + (3z_3-1)^2}{2}}
\ge \frac{1}{2} |3z_1-1-(3z_2-1)|.$ 
At the same time, 
$$(z_1+z_2)\sqrt{2}|2 \frac{z_1}{z_1+z_2}-1| = \frac{\sqrt{2}}{3} |3z_1-1-(3z_2-1)|.$$
To sum up, since \(\frac12-\frac{\sqrt2}{3}>0\), we have $g^{\mathcal{I}}_{2,CV}(z) \geq \left(\frac32-\sqrt2\right)|z_1-z_2|\ge 0.$

(2D) is violated in one way.  
From the inequality above, for the zero value to be attained we should necessarily have 
\(|z_1-z_2|=0\) and the second term is \(3z_3-1=0\), which immediately implies  \(z_1=z_2=z_3=\tfrac13\). Thus, most of the cases of subgroup means equality $z_3=\frac{z_1+z_2}{2}=\frac{1}{3}$ will be presented as cases with a  strictly positive between component.

Thus, just like in the population-share-weighting scheme, the within-group residual  satisfies (1B)-(1D) violating (1A). The performance of the between-group  residual appears better suited to this income-share-weighting scheme -- at least (2C) holds (it can be negative in the population-share-weighting scheme) and (2D) is violated in one way (in the population-share-weighting scheme (2D) was violated in two different ways). Thus, income-share-weighting appears to improve the behaviour of the between-group residual for the CV measure.

\subsection{Mean Log Deviation}
We have 
\begin{align*} g^{\mathcal{I}}_{1,MLD}(z) &=\frac{-\frac{1}{3}\left( \ln\left(3 z_1\right) + \ln\left(3 z_2\right) + \ln\left(3 z_3\right) \right)+\frac{1}{3}\left( \ln\left(\frac{3}{2} (z_1+z_2)\right) + \ln\left(\frac{3}{2} (z_1+z_2)\right) + \ln\left(3 z_3\right) \right)}{z_1+z_2} \\
&=
-\frac{1}{3\,(z_1+z_2)}\,
\ln\!\left( 4 \frac{z_1}{z_1+z_2} \cdot  \left(1-\frac{z_1}{z_1+z_2}\right)\right) \\ 
g^{\mathcal{I}}_{2,MLD}(z) & =-\frac{1}{3}\left( \ln\left(3 z_1\right) + \ln\left(3 z_2\right) + \ln\left(3 z_3\right) \right) +\frac{z_1+z_2}{2}\left( \ln\left(2 \frac{z_1}{z_1+z_2}\right) + \ln\left(2 \frac{z_2}{z_1+z_2}\right)\right)
\end{align*}

\begin{figure}[ht]
    \centering
    \begin{minipage}{0.47\textwidth}
        \centering
        \includegraphics[width=\linewidth]{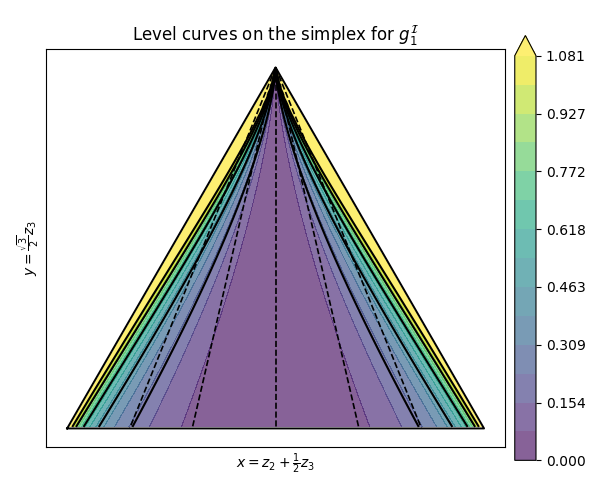}
    \end{minipage}
    \hfill
    \begin{minipage}{0.52\textwidth}
        \centering
        \includegraphics[width=\linewidth]{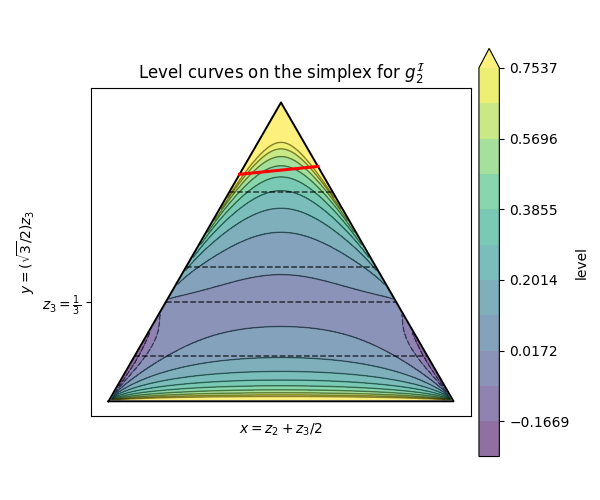}
    \end{minipage}
    \caption{\label{fig:MLD_DCI} Illustrations for MLD. Left: Level curves for $g^{\mathcal{I}}_{1,MLD}$ and rays constant at $\frac{z_1}{z_1+z_2}$. Right: Level curves for $g^{\mathcal{I}}_{2,MLD}$.}
\end{figure}

\paragraph*{Properties (1A)-(1D)}

Geometrically, as can be seen from the left-hand side plot in Figure \ref{fig:MLD_DCI}, level sets of $g^{\mathcal{I}}_{1,MLD}$ exhibit approximate but not exact invariance with the rays constant at $z_1/(z_1+z_2)$. Thus, although (1A) fails, it does so in a structured way. As confirmed by the same plot, (1B) holds.

(1C) holds because $ 4 \frac{z_1}{z_1+z_2} \cdot  \left(1-\frac{z_1}{z_1+z_2}\right)  \leq 4 \cdot \frac{1}{4}=1$, and, hence,  $-\ln \, \left(4 \frac{z_1}{z_1+z_2} \cdot  \left(1-\frac{z_1}{z_1+z_2}\right)\right) \geq 0$. 

(1D) holds too because $g^{\mathcal{I}}_{1,MLD}(z)=0$ iff  $ \frac{z_1}{z_1+z_2} \cdot \left(1-\frac{z_1}{z_1+z_2}\right)  = \frac{1}{4}$ which happens iff $\frac{z_1}{z_1+z_2}=1-\frac{z_1}{z_1+z_2}=\frac{1}{2}$, or, equivalently, $z_1=z_2$.

\paragraph*{Properties (2A)-(2D)} 

For properties of $g^{\mathcal{I}}_{2,MLD}$, the right panel of Figure \ref{fig:MLD_DCI} shows the extent to which level curves of $g^{\mathcal{I}}_{2,MLD}$ are not horizontal. 

The same plot visually confirms that (2B) does not hold. Along the red curve shown there, which is monotonic on the simplex,  the colours first change from lighter to darker and then, as $z_3$ continues to increase, change from darker to lighter.  

(2C) is violated as $g^{\mathcal{I}}_{2,MLD}$ can take negative values. 

(2D) holds partly, not fully. In other words, it is violated in one way. Namely, any $z$ with $z_3=\frac{1}{3}$ (hence, equal subgroup means $z_3=\frac{z_1+z_2}{2}$) will result in the zero value of $g^{\mathcal{I}}_{2,MLD}$. However, there will be other simplex points in the zero set of $g^{\mathcal{I}}_{2,MLD}$: namely, all points on the two symmetric curves described as 
$$(z_1,z_2,z_3) \in \Delta_2:\ z_3<\frac{1}{3},\ (z_1,z_2)=\left(
\frac{1-z_3 \pm \sqrt{(1-z_3)^2-4p(z_3)}}{2},
\frac{1-z_3 \mp \sqrt{(1-z_3)^2-4p(z_3)}}{2}
\right),$$
where 
$p(z_3)=\left(\frac{4^{\frac{3(1-z_3)}{2}}\,(1-z_3)^{-3(1-z_3)}}{27z_3}\right)^{\frac{2}{\,3z_3-1\,}}$. At \( z_3 = 0 \), these curves hit $(1,0,0)$ and $(0,1,0)$, and as \( z_3 \uparrow \frac{1}{3} \), they meet at
$\left( \frac{1}{3}, \frac{1}{3}, \frac{1}{3} \right)$.

MLD’s clean decomposition relies on population shares because its inequality concept is individual-centered rather than relational. Income-share weighting distorts the interpretation of MLD’s components.

Thus, from both MLD case here and the Theil index case in the population-share-weighting scheme we see the same pattern: the within component is approximately behaving as expected but the violations for the between components are particularly pronounced when the weighting scheme is chosen in an unsuitable way for a given index in this family. 

This highlights that the ``right'' notion of decomposability is probably measure-specific, not universal. 

The summary of violations is given in Table \ref{table:DCI}. 

\begin{table}[H]
\centering
\begin{threeparttable}
\caption{Income-share-weighted decomposability (DCI): summary of property failures}
\label{table:DCI}
\begin{tabular}{lcccccccc}
\toprule
Measure & (1A) & (1B) & (1C) & (1D) & (2A) & (2B) & (2C) & (2D) \\
\midrule
Theil & \checkmark & \checkmark & \checkmark & \checkmark & \checkmark & \checkmark & \checkmark & \checkmark \\
Gini  & partial     & \checkmark & \checkmark & \checkmark & \texttimes & \texttimes & \checkmark & Z$\Rightarrow$E fails \\
CV    & \texttimes  & \checkmark & \checkmark & \checkmark & \texttimes & \texttimes & \checkmark & Z$\Rightarrow$E fails \\
MLD   & $\approx$   & \checkmark & \checkmark & \checkmark & \texttimes & \texttimes & \texttimes & Z$\Rightarrow$E fails \\
\bottomrule
\end{tabular}
\begin{tablenotes}
\footnotesize
\item \textit{Notes:} 
Z$\Rightarrow$E means ``zero between-group component implies equal subgroup means.'' 
E$\Rightarrow$Z means ``equal subgroup means imply zero between-group component.'' 
``partial'' indicates region-specific validity; $\approx$ denotes structured but inexact invariance. 
\checkmark\ indicates the property holds; \texttimes\ indicates failure.
\end{tablenotes}
\end{threeparttable}
\end{table}

\section{Misinterpretation risks in applied work: stylised examples}
\label{sec:misinterpretation}

The geometric analysis of Sections \ref{sec:population}--\ref{sec:income} identified extents of 
violations of decomposability properties across measures and weighting schemes. 
This section translates failures of (2B), (2C), and (2D) into quantified numerical examples. Four scenarios are examined, each illustrating a qualitatively distinct mode of misinterpretation that arises when a practitioner applies a classical decomposition to a measure that does not satisfy the relevant decomposability property.

\begin{enumerate}
  \item[] \textbf{Scenario~1.} A {negative} between-group residual despite
        substantial mean differences (violation of (2C)). 
  \item[] \textbf{Scenario~2.} A {non-monotone} between-group component along
        a redistribution path that steadily widens subgroup mean differences (violation of (2B)).
  \item[] \textbf{Scenario~3.} A {zero} between-group component despite large
        mean differences (one direction of (2D) failure: $ Z \Rightarrow E$ fails).
  \item[] \textbf{Scenario~4.} A {positive} between-group component despite
       {equal} subgroup means (second direction of (2D) failure: $ E \Rightarrow Z$ fails).
\end{enumerate}

\noindent \textbf{Scenario 1: Negative between-group residual.} Table \ref{tab:risk1} traces the redistribution path $z(\lambda) = \left(0,\, \frac{2}{3}-\lambda,\, \frac{1}{3}+\lambda\right)$ for $\lambda \in [0, 0.40]$. At $\lambda = 0$ the subgroup means are equal. 
As $\lambda$ increases, income shifts from individual 2 to the singleton,  widening the mean gap to $\frac{3\lambda}{2}$.

\begin{table}[h]
\centering
\caption{Scenario~1 and 2: negative and non-monotonic between-group component for CV and Theil
         (population-share weighting) along the path
         $z(\lambda)=\left(0,\, \frac{2}{3}-\lambda,\, \frac{1}{3}+\lambda\right)$.}
\label{tab:risk1}
\smallskip
\begin{tabular}{ccccccc}
\toprule
$\lambda$ & $z_1$ & $z_2$ & $z_3$ & Mean gap & $g_{2,{CV}}$ & $g_{2,{Theil}}$ \\
\midrule
0.00 & 0.0000 & 0.6667 & 0.3333 & 0.0000 & +0.0572 & +0.0000 \\
0.05 & 0.0000 & 0.6167 & 0.3833 & 0.0750 & -0.0087 & -0.0292 \\
0.10 & 0.0000 & 0.5667 & 0.4333 & 0.1500 & -0.0540 & -0.0477 \\
0.15 & 0.0000 & 0.5167 & 0.4833 & 0.2250 & -0.0753 & -0.0561 \\
0.20 & 0.0000 & 0.4667 & 0.5333 & 0.3000 & -0.0710 & -0.0544 \\
0.25 & 0.0000 & 0.4167 & 0.5833 & 0.3750 & -0.0414 & -0.0427 \\
0.30 & 0.0000 & 0.3667 & 0.6333 & 0.4500 & +0.0111 & -0.0206 \\
0.40 & 0.0000 & 0.2667 & 0.7333 & 0.6000 & +0.1707 & +0.0566 \\
\bottomrule
\end{tabular}
\end{table}

At $\lambda=0$, $g_{2,{Theil}}=0$ and $g_{2,{CV}}>0$ (the CV between residual is non-zero at $\lambda=0$ due to the large within-subgroup inequality with $z_1=0$, which already contaminates the between residual). As the mean gap opens, both components  turn negative. $g_{2,{Theil}}$ remains negative until $\lambda \approx 0.35$, meaning the Theil between-group residual assigns a \emph{negative} contribution throughout a range where the singleton earns up to 60\% more per capita than the two-person subgroup. $g_{2,{CV}}$ similarly stays negative until $\lambda \approx 0.30$. With total income \$90{,}000, the configuration at $\lambda = 0.15$ corresponds to individual incomes
$(\$0,\$46{,}500,\$43{,}500)$,
the singleton earns 88\% more per capita than the two-person subgroup, yet both between-group residuals are negative.

\noindent \textbf{Scenario~2: Non-monotone between-group component.} This scenario has already been illustrated in Table 
\ref{tab:risk1} where we have $z_3>1/3$. Property (2B) then translates into requiring th eincreasing between-group residual as subgroup means {diverge} with increasing $\lambda$  (equivalently, with increasing $z_3$).  However, what we see is that as subgroup means {diverge},
the between-group residual component first declines up to  $\lambda \approx 0.15$ and then increases.  Thus, up to  $\lambda \approx 0.15$ the between-group residual 
signals declining inequality between groups while actual mean differences
are rising.

\noindent \textbf{Scenario~3: Zero between-group component despite large mean
differences.}
Table~\ref{tab:risk3} traces a path
$z(\lambda)=(\lambda,\,\tfrac{1}{2}-\lambda,\,\tfrac{1}{2})$ for
$\lambda\in[0,0.40]$, holding $z_{3}=0.5$ fixed.  As $\lambda$ increases,
individual 1's income grows at the expense of individual 2, while the
singleton's income share remains constant.  The subgroup means (individual 1 and 2 form one subgroup, and individual 3 forms the other subgroup) are invariant along this path and the two-to-one ratio of subgroup means never changes.

\begin{table}[h]
\centering
\caption{Scenario~3: between-group component for Gini
         (population-share weighting) for
         $z(\lambda)=\bigl(\lambda,\, \tfrac{1}{2}-\lambda,\, \tfrac{1}{2}\bigr)$.}
\label{tab:risk3}
\smallskip
\begin{tabular}{cccccc}
\hline
$\lambda$ & $z_{1}$ & $z_{2}$ & $z_{3}$ &
  Mean gap & $g_{2,\text{Gini}}$ \\
\hline
0.00 & 0.000 & 0.500 & 0.500 & 0.250 & $0.0000$ \\
0.05 & 0.050 & 0.450 & 0.500 & 0.250 & $0.0333$ \\
0.10 & 0.100 & 0.400 & 0.500 & 0.250 & $0.0667$ \\
\hline
\end{tabular}
\end{table}

 Table~\ref{tab:risk3} traces a path
 $z(\lambda)=\bigl(\lambda,\,\frac{1}{2}-\lambda,\,\frac{1}{2}\bigr)$
 for $\lambda\in[0,0.1]$, holding $z_3=0.5$ fixed.   At $\lambda=0$, i.e.\ $z=(0,\frac{1}{2},\frac{1}{2})$, the Gini
 between-group component is identically zero despite the singleton earning twice the subgroup mean.  The failure arises from a rank-based cancellation: when $z_1=0$ and $z_3=\max\{z_1,z_2\}=z_2$, the Gini's global rank distances produce exactly offsetting contributions that eliminate the between-group signal entirely. 

\noindent \textbf{Scenario~4: Positive between-group residual despite equal subgroup means.} 

 The risk is for a researcher to conclude that measurable inequality {between groups} exists when in fact it is entirely absent. Table~\ref{tab:risk4} evaluates two measures and considers the path $z(\lambda)=\bigl(\frac{1}{3}+\lambda,\,\frac{1}{3}-\lambda,\, \frac{1}{3}\bigr)$ parameterised by $\lambda\in[0,0.3]$.
 Along this path $z_3=\frac{1}{3}=(z_1+z_2)/2$ throughout, so the
 subgroup means are identically equal.

\noindent\begin{table}[h]
\centering
\caption{Scenario~4: positive between-group component despite equal
         subgroup means for Gini and CV (both population-share
         weighting).  Path: $z=\bigl(\frac{1}{3}+\lambda,\,
         \frac{1}{3}-\lambda,\,\frac{1}{3}\bigr)$.}
\label{tab:risk4}
\smallskip
\begin{tabular}{ccccccc}
\hline
$\lambda$ & $z_{1}$ & $z_{2}$ & $z_{3}$ &
  $g_{2,\text{Gini}}^{\phantom{I}}$ (DCP) &
  $g_{2,\text{CV}}^{\phantom{I}}$ (DCP) \\
\hline
0.000 & 0.333 & 0.333 & 0.333 & $0.0000$ & $0.0000$ \\
0.050 & 0.383 & 0.283 & 0.333 & $0.0167$ & $0.0086$ \\
0.100 & 0.433 & 0.233 & 0.333 & $0.0333$ & $0.0172$ \\
0.150 & 0.483 & 0.183 & 0.333 & $0.0500$ & $0.0257$ \\
0.200 & 0.533 & 0.133 & 0.333 & $0.0667$ & $0.0343$ \\
0.250 & 0.583 & 0.083 & 0.333 & $0.0833$ & $0.0429$ \\
0.300 & 0.633 & 0.033 & 0.333 & $0.1000$ & $0.0515$ \\
\hline
\end{tabular}
\end{table}

 At every row the between-group residual is strictly positive and growing for both measure-weighting pairs.  The values are analytically exact: $g_{2,{Gini}} = \lambda/3$ and $g_{2,{CV}} = \lambda(3-2\sqrt{2})$, confirming that both residuals are driven entirely by within-group heterogeneity (the growing gap between $z_1$ and $z_2$) rather than by any cross-group mean difference.


\section{Extension to other decomposability notions}
\label{sec:extensions}

The geometric framework developed in this paper is not tied to the two classical weighting schemes. Any alternative decomposability notion of the form
\[
I^3(z_1,z_2,z_3) = d(z_1,z_2)\,I^2(z_1,z_2) + d(z_3)\,I^1(z_3) + I^3\Bigl(\tfrac{z_1+z_2}{2},\tfrac{z_1+z_2}{2},z_3\Bigr)
\]
(with some weighting function $d(\cdot)$) can be analysed by defining the obvious analogues
\[
g_1^d(z) = \frac{I^3(z)-I^3(\bar z_1,\bar z_1,z_3)}{d(z_1,z_2)}, \qquad
g_2^d(z) = I^3(z) - d(z_1,z_2)\,I^2(z_1,z_2)
\]
and checking the eight properties (1A)–(1D), (2A)–(2D) on the simplex exactly as before.

A natural next step is to ask whether the same visual language is useful for \emph{non-classical} decompositions. Consider the recent axiomatic decomposition of the Gini coefficient proposed by \citet{HeikkuriSchief2026}. Their axioms (aggregation consistency across arbitrary partitions, path-independence, and a few others) deliver a unique within-group component that, in our three-person setting, reduces to
\[
f^{-1}\Bigl(\tfrac{2}{3}\bigr)^{\!1-\alpha} (z_1+z_2)^\alpha f\bigl(I^2(z_1,z_2)\bigr)
\]
for some strictly increasing function $f$ and some parameter $\alpha$.

The geometric implication of \citet{HeikkuriSchief2026} decomposition is  different from the classical case as the level sets of the within-group term are no longer required to coincide with rays of constant $z_1/(z_1+z_2)$. Instead, they are determined by the product $(z_1+z_2)^\alpha f(I^2(z_1,z_2))$. Plotting these level sets on the simplex therefore reveals at once how the new decomposition departs from population-share or income-share invariance  and whether the departure is economically attractive.

Imposing an additional axiom of conditional distribution independence uniquely characterizes the \citet{HeikkuriSchief2026} decomposition of the Gini coefficient for our three-person case:  the within-group component is
\[
G_W(z)
=
\frac{1}{3}|z_1-z_2|,
\]
and the corresponding between-group component is obtained residually as
\[
G_B(z)
=
\frac{1}{3}\left( |z_1-z_2|+|z_1-z_3|+|z_2-z_3| \right)
-
G_W(z) = \frac{1}{3}\left( |z_1-z_3|+|z_2-z_3| \right).
\]

From the perspective of the simplex diagnostics developed earlier, this decomposition satisfies neither population-share nor income-share invariance properties. Instead, it reflects a distinct organizing principle: within-group inequality depends on {absolute income gaps} inside the subgroup. The axioms of \cite{HeikkuriSchief2026} would require $G_W$ to satisfy our properties (1B)-(1D) (implications of their Axioms 1,4 and 5) in the  simple setting of 3 individuals (they require some other properties for more general partitions for larger populations) but not (1A). 
Thus, the \cite{HeikkuriSchief2026} axioms explicitly permit the within-group component/residual to depend on subgroup income shares. In the simplex, this implies that redistributions affecting $z_1+z_2$ may alter the within-group residual even when the relative split $z_1/(z_1+z_2)$ is unchanged. As a result, the natural reference geometry for this notion of decomposability is not given by the rays constant in $\frac{z_1}{z_1+z_2}$. \cite{HeikkuriSchief2026} Axiom 7 would require $G_B$ to take the value of zero when income distributions within subgroups are the same, which in our setting of $n=3$ and the split into $\{z_1,z_2\}$ and $\{z_3\}$ comes down to the case $z_1=z_2=z_3=\frac{1}{3}$. This is different from any of the properties (2A)-(2D) we considered earlier.   Axiom 8 in \cite{HeikkuriSchief2026} is not applicable to $G_B$ in the Gini setting since the overall Gini cannot be represented as function of $|z_1-z_2|/(z_1+z_2)$ alone.  

\begin{figure}[ht]
    \centering
    \begin{minipage}{0.47\textwidth}
        \centering
        \includegraphics[width=\linewidth]{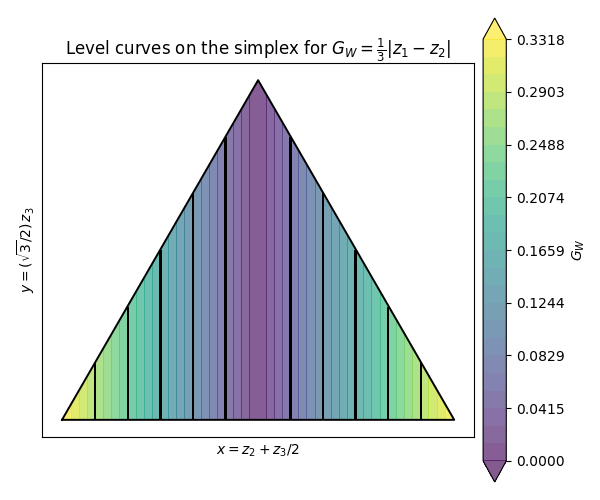}
    \end{minipage}
    \hfill
    \begin{minipage}{0.52\textwidth}
        \centering
        \includegraphics[width=\linewidth]{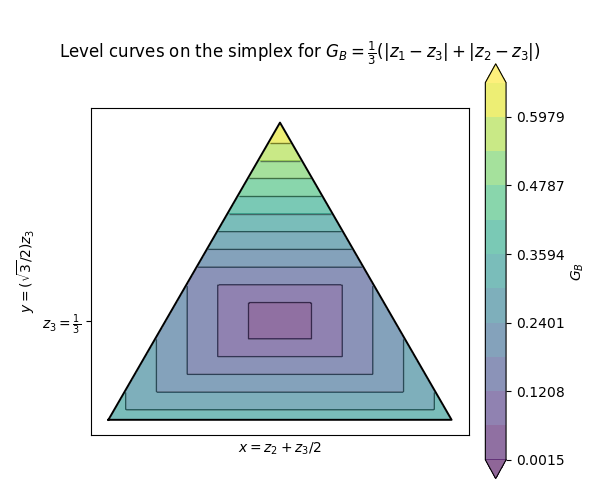}
    \end{minipage}
    \caption{\label{fig:GiniHS} Illustrations for Gini decomposition in \citet{HeikkuriSchief2026} Left: Level curves for $G_W$ Right: Level curves for $G_B$.}
\end{figure}

For completeness, Figure \ref{fig:GiniHS} shows the level curves of the within-group component/residual $G_W$ (on the left) and those of the between-group  component/residual $G_B$ (on the right), immediately confirming that neither classical invariance property holds for this decomposition. 

Different inequality measures  will have their  own aggregator functions $f$ and parameters $\alpha$ in the \cite{HeikkuriSchief2026} framework, ultimately making such decompositions measure-specific. The simplex analysis of this paper therefore can be seen as complementing the Heikkuri-Schief approach. It shows that their notion of decomposability corresponds to a different geometric invariance concept than the generic one considered in this paper.


\section{Conclusion}

This paper has analyzed decomposability in the simplest nontrivial setting of the three-person-population. 
It has shown that classical population-share-weighted and income-share-weighted decomposability can be made visually transparent by representing three-person income distributions on the income-share simplex. The resulting level-set and heat-map diagrams reveal exactly where and how the most commonly used inequality measures depart from the two classical notions of decomposability. 

The Gini coefficient and the coefficient of variation violate both notions, but the geometry localises the failures. For the Gini, the most serious violations occur precisely where rank reversals take place (the singleton income lies between the two subgroup incomes). For the coefficient of variation, the between-group residual can become negative  which a clear warning that CV may be ill-suited to subgroup analysis. The within-group residual  always preserves the expected Schur-convexity and non-negativity (property (1B) and (1C)) in all our examples. The between-group residual is where things go wrong most noticeably.

The figures in this paper are better viewed as analytical evidence rather than simple illustrations.
The results complement classical axiomatic characterizations by clarifying how decomposability fails when it does not hold. The central message of the paper is that   decomposability  is not a binary property and that different inequality measures violate decomposability in qualitatively distinct ways.

There are some opportunities for further research  that are beyond the scope of the present paper. One direction for future work could be to provide a scalar measure of the severity of decomposability violations. Such a metric could be based on deviation of the levels sets of $g_1$ and $g_2$ from the reference slices over the simplex. Such formals metrics could support formal comparisons across measures. A related strand of the applied literature, exemplified by \citet{CowellJenkins1995}, has approached the problem of non-decomposable measures from a different angle by constructing scalar summary statistics of how much of total inequality a given subgroup partition can account for, analogous to an $R^2$ in regression. The geometric framework developed in this paper can complement that approach.   Developing a formal bridge between the two approaches is naturally related to this direction for future work.

A second direction could be to provide an empirically weighted version of the simplex diagnostics, calibrated to the income distribution of a given dataset as in empirical work certain regions (those corresponding to moderate inequality with no extreme incomes) are far more commonly observed than others. The geometric framework developed here provides a natural foundation for each of them. 



{\normalsize  
\bibliographystyle{apalike}
\bibliography{overalldec}
}

\end{document}